\documentclass{article}
\usepackage{arxiv}
\usepackage[T1]{fontenc}
\usepackage{natbib}
\setcitestyle{authoryear,round,citesep={;},aysep={,},yysep={;}}

\renewcommand{\headeright}{Preprint}
\renewcommand{\undertitle}{Preprint}
\renewcommand{\shorttitle}{ProbeID}
\usepackage[hyphens]{url}
\usepackage{graphicx}
\usepackage{natbib}
\usepackage{caption}
\usepackage{algorithm}
\usepackage[noend]{algorithmic}
\usepackage{amsmath}
\usepackage{amssymb}
\usepackage{amsthm}
\usepackage{booktabs}
\usepackage{xcolor}
\usepackage{pifont}
\usepackage{tikz}
\usepackage{alphalph}
\usepackage{etoc}
\usepackage{multicol}
\usepackage{hyperref}
\let\probeidaddcontentsline\addcontentsline
\usetikzlibrary{positioning,arrows.meta,fit,backgrounds,calc,decorations.pathreplacing}
\definecolor{ink}{HTML}{1F2937}
\definecolor{trust}{HTML}{2A9D8F}
\definecolor{bad}{HTML}{E76F51}
\definecolor{method}{HTML}{3B5BA5}

\newtheorem{theorem}{Theorem}
\newtheorem{proposition}{Proposition}
\newtheorem{definition}{Definition}
\newcommand{\Dist}{\Delta}
\newcommand{\Disthat}{\widehat{\Delta}_m}
\renewcommand{\bot}{\textnormal{\textsc{abstain}}}

\title{ProbeID: Verifying MCP Tool Identity for\\ LLM Agents under Registry Drift}
\usepackage{authblk}

\author[1,*]{Guanxing~Chen}
\author[1]{Sining~Zhang}
\author[2]{Haohuai~He}
\author[3]{Qiujie~Lv}
\author[4,*]{Yu-An~Huang}
\author[1,*]{Zhi-An~Huang}
\affil[1]{City University of Hong Kong (Dongguan), Dongguan, China}
\affil[2]{The Hong Kong Polytechnic University, Hong Kong SAR, China}
\affil[3]{Zhengzhou University, Zhengzhou, China}
\affil[4]{Northwestern Polytechnical University, Xi'an, China}
\affil[*]{Corresponding authors: Guanxing Chen (\texttt{guanxing.chen@cityu-dg.edu.cn}), Yu-An Huang
(\texttt{yuanhuang@nwpu.edu.cn}) and Zhi-An Huang (\texttt{huang.za@cityu-dg.edu.cn})}

\begin{document}
\maketitle
\begin{abstract}
Agents select tools by copyable descriptions, which cannot establish which implementation runs. ProbeID compares canonicalized responses to schema-valid probes with a trusted admission reference and abstains on ambiguity, so a squatter must copy probe-observable behavior. With the gold present, independent probes and a fixed margin, standard concentration arguments give a shared probe budget logarithmic in the candidate count and bound the joint probability of a wrong decision. On cached MCP-IdentityDrift cases, strict choice reaches $0.956$ hit@1 while description routers fall near chance, but simpler exact comparisons tie it, MMD better tolerates heavy value noise, and a $32$B behavioral judge is similarly accurate at higher cost. Random probes from agent banks and a prospective held-out study of $38$ tools test theory-set thresholds. On the agent banks, equality refuses every reachable replacement from $64$ probes and the bounded distance from $512$, and sharper post hoc gates halve both. Two tools whose drift no bank probe reaches always pass. In an MCP-Bench agent, verification cuts tainted tasks from $15$ to $9$ of $16$ at unchanged task success. Over eleven weeks, $7$ of $59$ real tools changed responses despite version pins. Verification depends on reference integrity, probe coverage, and response noise.
\end{abstract}

\etocdepthtag{main}

\section{Introduction}
\begin{figure}[t]
\centering
\includegraphics[width=\textwidth]{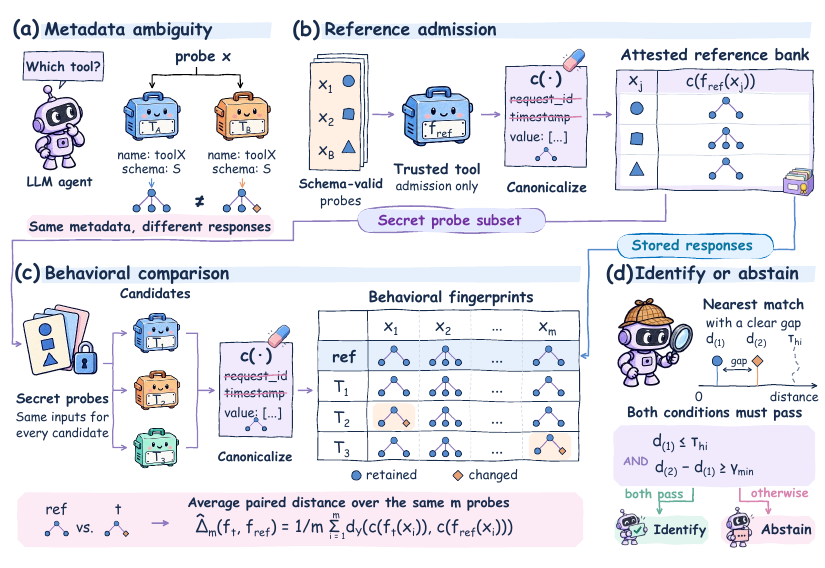}
\caption{\textbf{Behavior decides what copied metadata cannot.} (a) Identical tool metadata can hide different responses. (b) Admission stores canonicalized trusted responses to schema-valid probes (M1). (c) Selection sends secret bank probes to candidates and compares their paired responses with the stored reference (M2 to M4). (d) Multiple candidates pass the distance and gap gates (M5), while a sole candidate faces only the distance gate. ProbeID otherwise abstains (M6). Glyphs, names, and distances are schematic.}
\label{fig:teaser}
\end{figure}

Large language model (LLM) agents increasingly act through external tools, and the Model Context Protocol (MCP) has turned this into a fast-growing open ecosystem of community-published tool servers \citep{modelcontextprotocol,beyondmcp2506}. An agent that needs a capability searches a large registry and ranks tools by their advertised descriptions \citep{qin2023toolllm,mcpzero2506}, so unauthenticated, attacker-editable text decides which implementation runs \citep{sneh2025tooltweak}. Audits find most tool descriptions underspecified \citep{smelly2602}, server behavior departing from the descriptions at registry scale \citep{li2026misleadingdescriptions}, and squatting among the main attack vectors \citep{typosquat2601}. The protocol's security interest group lists runtime drift and tool identity as open problems \citep{modelcontextprotocol2026security}.

However, description routing and admission-time binding both assume that declared metadata fixes behavior, and registry drift breaks that assumption. A squatter copies the description of a trusted tool but implements different behavior \citep{beurerkellner2025toolpoisoning,mcptox2508}, so a description router admits it and may select it. Admission controls bind definitions or provenance \citep{etdi2506,ztrsquat2504,metere2026attested}, but a definition bound at admission does not expose a later change in behavior. In an adaptive-distractor setting, certified correctness of metadata-based selection falls to about twenty percent after ten rounds \citep{toolcert2510}. No description matcher can recover the intended tool from identical descriptions.

To remove the dependence on declared metadata, we identify a tool by its behavior, as black-box model fingerprinting identifies a model \citep{pasquini2025llmmap,gao2025modelequality}. Behavioral identity here means agreement of probe-observable responses with a reference recorded at admission, not endpoint or cryptographic identity. Probing candidates at selection time and comparing what they return forces a squatter to match behavior rather than text. The comparison observes only responses, so the probes must reach the difference and the reference must be trusted (Figure~\ref{fig:teaser}).

ProbeID is a post-admission behavioral verifier that generates schema-driven probes, canonicalizes the responses, and selects the candidate nearest the reference under a bounded behavioral distance, abstaining when no candidate matches or the decision is ambiguous. At a fixed margin, adding $k$ squatters raises the number of shared probe inputs only by a logarithmic factor, while a router the squatters mimic falls to chance $1/(k+1)$. We validate the protocol on MCP-IdentityDrift, built from real cached tool behavior \citep{guo2024stabletoolbench} with injected schema-compatible squatters, and test its error bounds with random probes from the attested banks of an agent study and from a prospective held-out study. The contribution is the protocol, which does not depend on one comparison. Exact equality ties the bounded distance on the cached cases and verifies deterministic drift with fewer probes, and a set-level maximum mean discrepancy (MMD) tolerates heavy value noise better.

Our contributions are as follows.
\begin{itemize}
\item A reference-based verification protocol that separates identity selection from abstention (Algorithm~\ref{alg:probeid}) and calibrates both with standard concentration arguments. For independent probes it needs a shared budget $m=O(\gamma^{-2}\log((n+k)/\delta))$ and, with the gold present, keeps the joint probability of a wrong decision below $\delta$, with a matching lower bound on the leading term under independent candidate channels (Appendices~\ref{app:proof} to~\ref{app:budgetbounds}).
\item MCP-IdentityDrift and an evaluation that separates operating points (Table~\ref{tab:operating}). Routers approach $1/(k+1)$ under copied-description rug pull while ProbeID reaches $0.956$ strict-choice hit@1. Random probes from the agent banks and from a prospective held-out study show where theory-set thresholds refuse replacements and where their resolution ends (Result~2). An agent gate cuts tasks tainted by the rug pull from $15$ to $9$ of $16$, a harm that task success does not register (Result~4).
\item A census of $16$ real published MCP servers re-probed eleven times over eleven weeks. Canonicalization succeeds on every recorded response, most tools are deterministic, and $7$ of $59$ depart from their recorded responses under unchanged version pins, three of them after a reinstall at unchanged versions (Appendix~\ref{app:census}).
\end{itemize}

\section{Related work}
\paragraph{Tool retrieval and selection.} Tool selection is usually cast as retrieval over descriptions. ToolLLM trains a dense retriever \citep{qin2023toolllm} and MCP-Zero discovers tools from server metadata \citep{mcpzero2506}. Bridge models recover part of what dense retrieval loses under vague instructions \citep{trb2604}, rerankers over tool cards add little \citep{shi2025toolret}, and documentation augmentation improves descriptions \citep{play2prompt2503}. All select using descriptions, so a copied description inherits its retrieval score.

\paragraph{Tool ecosystem security.} Squatting, description tampering and rug pulls are documented in the MCP ecosystem, whose specification treats self-declared tool annotations as untrusted \citep{modelcontextprotocol,typosquat2601,beyondmcp2506}. Disclosures, taxonomies and benchmarks report tool poisoning, rug pulls and message history exfiltration in deployed servers \citep{beurerkellner2025toolpoisoning,kosman2025poisoneverywhere,beurerkellner2025whatsappmcp,beyondmcp2506,mcptox2508}. Admission defenses bind immutable definitions, registry trust or provenance \citep{etdi2506,ztrsquat2504,metere2026attested,mas2026registry}, and runtime detectors screen for deviation from declared intent or for injected content \citep{huang2026connor,zhou2026mcpshield,xing2025mcpguard}, without a trusted reference or selection among squatters. ProbeID complements these defenses at use time. It detects an admitted rug pull when the trusted reference and the drawn probes expose the change, and it cannot see exfiltration that leaves responses unchanged.

\paragraph{Certification of tool selection.} LLMCert-T, the closest formal neighbor, certifies a tool selection pipeline over its metadata and finds certified correctness collapsing under adaptive distractors \citep{toolcert2510}. ProbeID instead gives a conditional guarantee from probe-observable behavior under an explicit margin model, and composing it with a certificate over metadata is left open.

\paragraph{Identity from behavior.} Examining an artifact itself rather than its declared metadata recurs across conformance and differential testing, query learning, stateful API fuzzing, network service scanning, software birthmarks, and black-box model fingerprinting \citep{chow1978fsm,mckeeman1998differential,angluin1987lstar,atlidakis2019restler,lyon2009vscan,tamada2004birthmark,myles2005kgram,podgurski1993behavior,pasquini2025llmmap,shao2026zeroprint,gao2025modelequality,zhu2026rank,karvonen2025difr,zhang2026spoofing}. Model Equality Testing uses a two-sample test to check whether an API still serves a trusted model, and LLMmap identifies which of a known set of models produced a set of traces, so selection among candidates alone does not distinguish ProbeID. Here the candidates are adversarially authored to share a schema and a description. ProbeID pairs the margin the probes must reach with a budget bound, a matching lower bound on its leading term, and an abstention rule with a joint error bound (Appendix~\ref{app:relwork}).

\paragraph{Concurrent work.} Concurrent work generates schema-valid probes for MCP servers to adjudicate execution risk rather than identity \citep{flowguard2607} and measures how far an agent skill's behavior departs from its declared capability \citep{behavintegrity2605}. Other work tracks LLM endpoint stability and agent workflow regressions with behavioral fingerprints \citep{leshin2026endpointstability,bhardwaj2026agentassay}, and TrustShiftProbe audits live payloads against baselines learned while a server was trusted \citep{rostamzadeh2026trustshift}. Each checks one artifact against its own declaration or history, whereas ProbeID probes all candidates on inputs with recorded trusted responses and selects among them.

\section{Behavioral tool identity}

\subsection{Problem setup}
An LLM agent receives a user intent $q$ and a registry $R$ of $n$ tools, to which drift can add up to $k$ squatters. A tool $t$ has a description $\mathrm{desc}(t)$, a typed schema $S_t$ defining the valid inputs $X_t$, and a behavior $f_t:X_t\to Y_t$, a partial and possibly stochastic map returning an output, a status code and a state delta. A description router forms candidates $C(q)\subseteq R$ by lexical or semantic similarity.

Let $t^\star$ be the gold tool whose behavior $q$ requires. Registry drift introduces schema-compatible squatters that copy or approximate $\mathrm{desc}(t^\star)$ with different behavior, description edits, and name or schema mimicry. The task is behavioral identity recovery, which returns a single $\hat c\in C(q)$ behaviorally equivalent to $t^\star$, or \bot{} when no candidate is confidently identifiable. Two tools are behaviorally equivalent on a probe distribution $D$ over schema-valid inputs when their canonicalized responses agree for $D$-almost every probe on the observed channels. A decision incurs the $0/1$ identity loss $L_{\mathrm{id}}(\hat c)=\mathbf{1}[\hat c \text{ not equivalent to } t^\star]$, and an abstention is counted separately as lost coverage. This notion is weaker than endpoint identity and cannot see side effects that never reach a response, behavior outside the probe support, or a reference recorded from a compromised tool. A tool that exfiltrates message history \citep{beurerkellner2025whatsappmcp} while preserving every probe-visible response ties the gold when both are candidates and passes the single-candidate gate when it replaces the gold.

\subsection{ProbeID}
ProbeID is a post-admission runtime verifier. At admission each tool's trusted reference behavior $f_{\mathrm{ref}}$ is recorded as an attested bank of $B$ schema-valid probes with their canonicalized responses. At selection ProbeID secretly subsamples $m$ bank probes and probes only the current candidates, so no live reference endpoint is needed.

\paragraph{Schema-driven probes (M1).} ProbeID generates the bank at admission by type-aware enumeration over boundary values, enums, and required or optional fields (Appendix~\ref{app:probegen}), and selection draws a secret subsample $\{x_1,\dots,x_m\}$. Theorem~\ref{thm:bound} models probes as i.i.d. draws from $D$. The same bound holds for uniform sampling without replacement from the attested bank with $D$ read as the bank distribution, and a population margin transfers to the bank only if the bank was itself sampled from that population (Appendix~\ref{app:proof}). Pure deterministic enumeration is a heuristic outside the guarantee, and heterogeneous or aliased schemas are filtered upstream.

\paragraph{Canonicalization and bounded distance (M2, M3).} A canonicalizer $c(\cdot)$ drops named volatile fields such as timestamps and request ids and sorts mapping keys. A bounded distance $d_Y\in[0,1]$ then scores structural disagreement over the response tree, with status codes and state deltas as further channels combined by a probabilistic union (Appendix~\ref{app:dy}). The analysis does not need this distance to dominate other paired comparisons.

\paragraph{Behavioral distance and identity rule (M4, M5).} For the probe distribution $D$, the behavioral distance between two tools is
\begin{equation}
\Dist(f,g) = \mathbb{E}_{x\sim D}\big[\, d_Y(c(f(x)), c(g(x))) \,\big].
\label{eq:dist}
\end{equation}
$\Dist$ is bounded and symmetric, and the analysis needs only these properties and the margin, not a triangle inequality. Deterministic tools have $\Dist(f,f)=0$. Stochastic tools may have positive self distance $\Dist(f,f)=\mu_{t^\star}\ge 0$ between two independent responses to the same probe. The empirical estimate over $m$ shared probes is $\Disthat(f,g)=\frac1m\sum_{i=1}^m d_Y(c(f(x_i)),c(g(x_i)))$, the behavioral fingerprint of $t$ is $\phi(t)=(c(f_t(x_1)),\dots,c(f_t(x_m)))$, and given a trusted reference $f_{\mathrm{ref}}$ the nearest-fingerprint rule selects
\begin{equation}
\hat c = \arg\min_{t\in C(q)} \Disthat(f_t, f_{\mathrm{ref}}).
\label{eq:nearest}
\end{equation}

\paragraph{Abstention and escalation (M6).} ProbeID returns \bot{} when the smallest distance exceeds $\tau_{\mathrm{hi}}$, so no candidate matches the reference, or when the gap to the second nearest is below $\gamma_{\min}$, so identity is ambiguous (Algorithm~\ref{alg:probeid}). The caller may then escalate to confirmation or to the description router, whose choice does not inherit the behavioral error bound. A single registered candidate, such as an endpoint a squatter replaced, faces only the absolute gate (Appendix~\ref{app:abstain}).

\begin{algorithm}[t]
\caption{ProbeID identity verification}
\label{alg:probeid}
\begin{algorithmic}[1]
\STATE \textbf{Input:} candidates $C(q)$, attested bank $\{(x_j, c(f_{\mathrm{ref}}(x_j)))\}_{j=1}^{B}$, budget $m$, thresholds $\tau_{\mathrm{hi}},\gamma_{\min}$
\STATE secretly subsample $x_1,\dots,x_m$ with their bank reference $\phi(\mathrm{ref})$ \hfill (M1)
\FOR{each $t\in C(q)$}
  \STATE $\phi(t)\gets (c(f_t(x_1)),\dots,c(f_t(x_m)))$ and $d_t \gets \Disthat(f_t,f_{\mathrm{ref}})$ \hfill (M2 to M4)
\ENDFOR
\STATE sort $d_{(1)}\le d_{(2)}\le\cdots$
\IF{$|C(q)|=1$}
  \STATE \textbf{return} the sole candidate if $d_{(1)}\le\tau_{\mathrm{hi}}$, else \bot{}
\ENDIF
\IF{$d_{(1)} > \tau_{\mathrm{hi}}$ \OR $d_{(2)}-d_{(1)} < \gamma_{\min}$}
  \STATE \textbf{return} \bot{} \hfill (M6)
\ENDIF
\STATE \textbf{return} $\arg\min_t d_t$ \hfill (M5)
\end{algorithmic}
\end{algorithm}

\section{Identity guarantee under squatting}
\label{sec:theory}
We now state when the nearest-fingerprint rule in Equation~\eqref{eq:nearest} is correct. The key quantity is a behavioral separation margin.

\begin{definition}[Margin]
Write $\mu_t=\Dist(f_t,f_{\mathrm{ref}})$ for the expected behavioral distance of candidate $t$ to the reference, the expectation taken over both the probes $x\sim D$ and any tool randomness. A candidate set has margin $\gamma$ if every behaviorally wrong candidate $w$ is separated from the gold in mean, $\mu_w-\mu_{t^\star}\ge\gamma$ on $D$.
\end{definition}

The margin depends on the probe distribution. For a deterministic gold that matches the reference pointwise, $\gamma$ is positive exactly when every squatter has positive distance on a set of positive probe mass (Appendix~\ref{app:sep}). A squatter at zero distance on $D$ is equivalent to the gold under this comparison, and any harm it does lies outside the probe support. A faithful deterministic gold has self distance $0.00$, as on the deterministic census tools (Appendix~\ref{app:census}), and the variance $\sigma^2$ covers sampled inputs and response randomness. Theorem~\ref{thm:bound} assumes observations independent across probes, so stateful tools lie outside it and are treated empirically (Appendix~\ref{app:statefulall}).

\begin{theorem}[Probe complexity]
\label{thm:bound}
Let $d_Y\in[0,1]$, and suppose the candidate pool contains the gold tool $t^\star$ among at most $n+k$ tools with margin $\gamma>0$. Draw $m$ probes i.i.d. from $D$, with per-probe observations independent given the probes. For any $\delta\in(0,1)$, if
\begin{equation}
m \;>\; \frac{2}{\gamma^2}\,\log\!\frac{2(n+k)}{\delta},
\label{eq:bound}
\end{equation}
then the nearest-fingerprint rule returns a behaviorally correct tool with probability at least $1-\delta$.
\end{theorem}

The proof combines Hoeffding's inequality \citep{hoeffding1963probability} with a finite-class union bound \citep{shalevshwartz2014understanding} (Appendix~\ref{app:proof}), and the gap gate below is the stopping rule of action elimination \citep{evendar2006action}. For a pool with the gold and at least two candidates, Algorithm~\ref{alg:probeid} uses $\gamma_{\min}=2\epsilon$, where $\epsilon=\sqrt{\log(2(n+k)/\delta)/(2m)}$, and $\tau_{\mathrm{hi}}$ at least $\epsilon$ above the gold self distance. Under the stated sampling conditions, the probability of deciding and being wrong is at most $\delta$, and the rule returns a correct candidate when $\gamma\ge4\epsilon$ and no twin of the gold is present (Theorem~\ref{thm:abstain}). The bound is on the joint event, and among decided cases the error is bounded only by $\delta/P(\text{decide})$, so an observed error rate among decisions is not a $\delta$-level certificate. A sole candidate faces only the absolute gate, and with $\tau_{\mathrm{hi}}$ exactly $\epsilon$ above the gold self distance it refuses a replacement with probability at least $1-\delta$ when its margin exceeds $2\epsilon$ (Appendix~\ref{app:abstain}).

At a fixed margin, $k$ schema-compatible squatters affect the shared-probe budget through $\log((n+k)/\delta)$. Querying every candidate still requires up to $m(n+k)$ candidate tool calls, besides the one-time admission calls that record the bank.

\paragraph{Variation and the $\sigma^2$ factor.} Hoeffding gives the worst-case $1/\gamma^2$ rate. When the per-probe distance has variance at most $\sigma^2$, a Bernstein bound gives $m>\big(8\sigma^2\gamma^{-2}+\tfrac43\gamma^{-1}\big)\log(2(n+k)/\delta)$ (Appendix~\ref{app:bernstein}). For fixed target error and independent candidate channels, a two-point and Fano construction give an $\Omega(\sigma^2\gamma^{-2}\log(n+k))$ lower bound on the leading term (Appendix~\ref{app:lower}).

\section{Experiments}
The experiments ask whether behavior resolves copied descriptions, whether budgets and error rates follow the theory under random probes, which comparison suits which noise, and how far verification reduces corrupted agent inputs.

\subsection{Setup}
We build MCP-IdentityDrift on the StableToolBench virtual API server, whose cache returns deterministic outputs for cached arguments \citep{guo2024stabletoolbench}, over the ToolBench corpus \citep{qin2023toolllm}. A recursive loader yields 3493 tools with at least four structured cached responses. A gold tool's cached arguments form its probe set and its cached responses the trusted reference $f_{\mathrm{ref}}$, and $k$ schema-compatible squatters keep its description while mutating output values or silently dropping a field, a rug pull. Identity runs use up to 12 probes and three registry seeds with adversarial tie-breaking, so a behavioral twin counts as a miss. A gold enters only if some squatter is separable from it, so hit@1 is conditional on that filter (Appendix~\ref{app:impl} counts the excluded golds). Verification runs on CPU, and headline comparisons report $95\%$ paired bootstrap intervals over $10^4$ resamples. Baselines run on the 402 tools with a real ToolBench description, 150 gold per seed. BM25 \citep{qin2023toolllm}, a dense retriever in the ToolRetriever style, and an LLM reranker serving Qwen2.5-7B over tool cards \citep{sun2023rankgpt,shi2025toolret} share candidates and backend, and a random selector and an oracle give the $1/(k+1)$ floor and the ceiling.

\subsection{Result 1: identity under registry drift}
Copied descriptions remove the identity signal available to every description router (Table~\ref{tab:main}), and definition hash pinning is blind to the same behavior-only change \citep{aisdk2026tooldrift}. These controls isolate information access without showing superiority over other behavioral methods. On a separate clean cohort the lightweight behavioral comparisons tie (Appendix~\ref{app:baselines}), exact equality needs fewer probes on deterministic agent tools, and MMD better tolerates heavier value noise. The informative comparisons are therefore among behavioral methods, and their order depends on the noise regime and the budget. Appendix~\ref{app:ksweep} extends the candidate sweep.

Figure~\ref{fig:main}(a) adds the judge cohort, $300$ gold per seed against the headline's $150$. The judges' prompts cut each serialized response to $200$ characters, whereas ProbeID compares complete canonicalized responses. On complete-response prompts that fit, the 32B judge's accuracy changes by at most $0.006$, but $8$ to $17\%$ of complete prompts exceed its $32$k-token context (Appendix~\ref{app:judge}). Qwen2.5-7B trails at every $k$, while Qwen2.5-32B is ahead of ProbeID at $k\in\{2,4\}$, where strict tie-breaking scores ProbeID's ties as misses, and behind at $k\in\{1,8\}$. The 32B judge's decision step takes $3.40$ seconds on a dedicated $46$GB GPU against $3.527$ milliseconds for ProbeID on one CPU core. Both timings exclude probe calls and use different hardware, so they establish neither end-to-end acceleration nor a reproducibility advantage (Appendix~\ref{app:determinism}). At $k{=}8$ the gap over the better router is $0.844$, with a $95\%$ gold-clustered bootstrap interval of $[0.807, 0.880]$ over $307$ distinct gold tools. The $4.4\%$ strict-choice misses at $k\ge2$ are mutations that canonicalized back to the gold on every cached probe, realized twins that strict tie-breaking scores as misses (Appendix~\ref{app:riskcov}).

\begin{table}[t]
\centering
\caption{\textbf{Operating points kept apart.} Decided and wrong are fractions of all checks. With the gold present, wrong is the joint rate Theorem~\ref{thm:abstain} bounds, and with the gold absent it is the acceptance of a replacement. Cached rows use the fixed cache-order probes of the $450$ headline cases at $k{=}8$, which meet no sampling condition, and strict choice scores a tie as wrong. Bank rows draw $m$ probes at random from the agent study's banks, $1{,}000$ draws per tool, with theory-set thresholds at $\delta{=}0.05$ (Appendix~\ref{app:banksample}). The two unreachable tools match the gold on every bank probe but corrupt agent calls. A check costs $\min(m,B)$ probe calls per candidate.}
\label{tab:operating}
\small
\setlength{\tabcolsep}{4pt}
\begin{tabular}{@{}lllcc@{}}
\toprule
evidence & probes & rule & decided & wrong \\
\midrule
cached cases, gold present & fixed, $\le12$ & strict choice & 1.000 & 0.044 \\
cached cases, gold present & fixed, $\le12$ & per-case radius & 0.153 & 0.000 \\
banks, gold present, 8 reachable tools & random, $m{=}32$ & Algorithm~\ref{alg:probeid}, equality & 0.784 & 0.000 \\
banks, gold absent, 8 reachable tools & random, $m{=}32$ & single gate, equality & 0.006 & 0.006 \\
banks, gold absent, 2 unreachable tools & random, any $m$ & single gate, either & 1.000 & 1.000 \\
banks, 77 golds and 67 unaffected tools & random, any $m$ & single gate, either & 1.000 & 0.000 \\
\bottomrule
\end{tabular}
\end{table}

ProbeID stays at $0.956$ as a squatter's description diverges (Appendix~\ref{app:descsim}). On two real JSON serializers the small margin constrains the budget, with no misidentification (Appendix~\ref{app:jsonimpl}), and among near-duplicate vendor descriptions a dense router cannot identify the recorded sibling (Appendix~\ref{app:natural}).

\subsection{Result 2: probe budget and error control}
\begin{figure}[t]
\centering
\includegraphics[width=\textwidth]{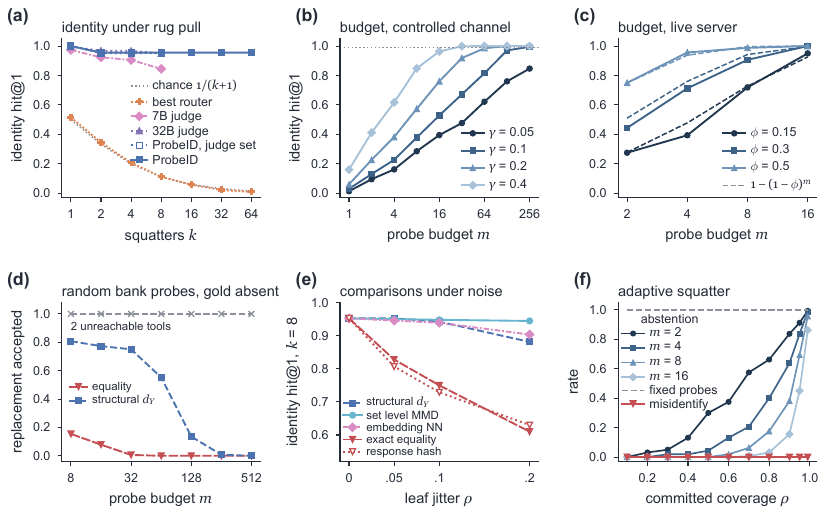}
\caption{\textbf{Identity and abstention across operating regimes.} (a) Strict-choice hit@1 on the headline and judge cohorts, with routers near chance under copied descriptions. (b) Probe budget against margin. (c) Fresh live probes against a near twin, with the sampling prediction dashed. (d) Acceptance of a rug-pull replacement, gold absent, under random bank probes and theory-set thresholds, averaged over the eight reachable tools. No draw refuses a gold. (e) Comparison methods under retained value noise. (f) Identification and abstention against a committed coverage squatter.}
\label{fig:main}
\end{figure}

The deterministic cache does not exercise the noisy regime that $16$ of the $59$ census tools show (Appendix~\ref{app:census}). In a Bernoulli mismatch channel calibrated to cached behavior, where a squatter diverges on a $\gamma$ fraction of probes and tool noise sets a positive self distance and variance $\sigma^2$, hit@1 rises with $m$ (Figure~\ref{fig:main}(b), Appendix~\ref{app:realbudget}).

The smallest budget $m^\star$ reaching $0.99$ hit@1 keeps $m^\star\gamma^2$ near constant over a fourfold range of $\gamma$ on a doubling grid. The product grows from the $2.6$ to $5.8$ band at a gold self distance of $0.18$ to the $5.1$ to $10.2$ band at $0.34$, the direction the $\sigma^2$ factor predicts (Table~\ref{tab:bound}). Against a live near twin that diverges on a $\phi$ fraction of inputs, measured hit@1 tracks $1-(1-\phi)^m$ (Figure~\ref{fig:main}(c)). Live servers with byte-identical tool cards are identified at $1.00$ by eight probes where a description router sits near chance (Appendix~\ref{app:live}).

The benchmark's fixed cache-order probes do not meet the theorem's sampling condition, so we test the bounds on the attested banks of the agent study (Result~4), which recorded every entry, up to $512$ probes, against the gold twice and its rug-pull replacement once. We draw $m$ probes uniformly without replacement from each of the $77$ banks, $1{,}000$ draws per tool and budget, with every threshold fixed by the theory at $\delta{=}0.05$ (Table~\ref{tab:operating}, Figure~\ref{fig:main}(d), Appendix~\ref{app:banksample}). No observed rate violates its bound in the $1{,}282$ cells where one applies, which checks threshold implementation rather than tool behavior, since a bank mean is an exact expectation under its own sampling (Table~\ref{tab:bankcheck}). The informative quantity is refusal. With the gold absent, equality accepts a replacement of the eight reachable drifted tools at a mean rate of $0.154$ at $m{=}8$, $0.006$ at $m{=}32$ and $0$ from $m{=}64$. The structural distance needs $512$ probes, because on the conversion and calculator tools equality's margin is four to five times larger. Neither comparison refuses a gold or any of the $67$ unaffected tools, and the two tools whose drift no bank probe reaches pass at every budget, the zero-margin case the theorem excludes. With the gold present no draw returns the squatter, which a deterministic gold at distance zero guarantees. A prospective held-out study repeats the test on the $38$ read-only tools of eight census servers with seven replacement policies, from full and partial value drift to legitimate updates, and draws fresh random banks only after fixing its rules and policies (Appendix~\ref{app:heldout}). At the full bank every candidate with a bank margin above $0.2$ is refused and no gold is, but drift on one input in twenty passes on $14$ of $22$ local tools, because the worst-case radius cannot resolve so small a margin. An exact hypergeometric test for equality and a finite-population radius for the structural distance, added once these results were known, halve the observed refusal budgets on reachable agent replacements, from $64$ to $32$ probes and from $512$ to $256$. At the full bank they refuse every held-out replacement whose bank distance exceeds the admission self distance, and the structural gate also refuses every answer-changing legitimate update tested (Appendix~\ref{app:exact}). A sequential structural gate (Proposition~\ref{prop:sequential}) refuses full value drift on local tools after $8$ probes on average instead of $512$.

\subsection{Result 3: complementarity, and drift in real registries}
ProbeID is a verifier, not a replacement for routing. It abstains on a constructed twin family whose members differ only in description, where a lexical router recovers the gold (Appendix~\ref{app:twin}). On $40$ identity trials over $20$ census tools in $10$ schema-compatible sibling groups with distinct genuine descriptions, a dense router scores $1.00$ and ProbeID $0.90$, abstaining without misidentification on one pair at distance $0$, so verification follows routing.

Canonicalization zeroes the gold self distance under the volatile fields it names but can hide drift confined to them (Appendix~\ref{app:canon}). Under residual noise, $0/1$ equality and hash baselines degrade most and a set-level MMD performs best under heavier value jitter (Figure~\ref{fig:main}(e)). On the agent's deterministic tools, however, the $0/1$ test verifies with fewer probes (Appendices~\ref{app:baselines} and~\ref{app:agent}). At $\rho{=}0.2$ and $k{=}8$, the MMD advantage is $0.063$, with a pointwise $95\%$ gold-clustered interval of $[0.043,0.085]$ (Appendix~\ref{app:baselines}). No single comparison is best in every regime.

\paragraph{Drift in the wild.} On the $16$ census servers, eleven byte-for-byte replays of the recorded probes over eleven weeks show $7$ of $59$ tools moving beyond their own $t_0$ noise, setting aside three measurement artifacts. Three weather tools reflect legitimate content updates, and an AWS documentation recommender climbs from $0.10$ to $1.00$ by $55$ days at an unchanged recorded version while two sibling tools stay at $0.00$. Three further tools moved after a reinstall at unchanged versions, because their SDK declares its validation library as a range spanning two major versions and the validation errors changed form (Figure~\ref{fig:agentview}(a)). Version pins therefore do not fix responses, and none of these changes is evidence of malice (Figure~\ref{fig:agentview}(b), Appendix~\ref{app:census}).

\subsection{Result 4: verification in an agent}
\begin{figure}[t]
\centering
\includegraphics[width=\textwidth]{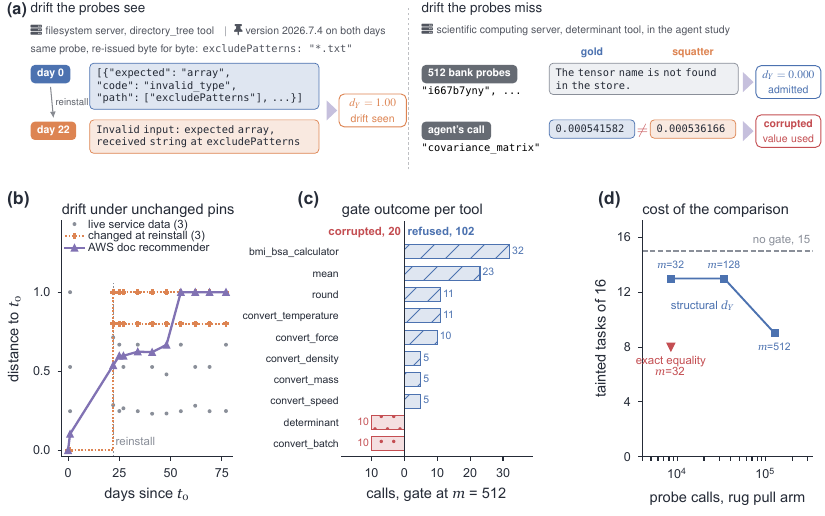}
\caption{\textbf{Visible and missed drift.} (a) A changed validation error after reinstall is visible, while bank probes miss a tensor value changed on an agent call. (b) Per-tool distance from admission over eleven weeks. (c) Refused and corrupted agent calls by tool. (d) Tainted tasks and probe calls under structural distance and exact equality, one run each.}
\label{fig:agentview}
\end{figure}

We test whether verification changes the tool values an agent receives. On MCP-Bench \citep{mcpbench2508} we run the $16$ tasks that touch four servers for which we hold a gold and a squatter, benign or rug pulled, with and without the gate, at one seed under greedy decoding. Gold replay, rather than benchmark scores, defines the outcome. Every call to a server with a gold is re-issued to it with identical arguments, so a difference is a value the agent should not have been given. Unverified, the rug pull corrupts $93$ of $1476$ adjudicable calls and taints $15$ of $16$ tasks. With the gate drawing each tool's whole bank of up to $512$ probes, it corrupts $20$ of $1159$ adjudicable calls and taints $9$ tasks, with $102$ calls refused instead (Figure~\ref{fig:agentview}(c)). Both benign arms have no corrupted call, and the gated benign arm refuses none of its $269$ verdicts. All three end-to-end runs of each arm taint equally many tasks, but corrupted-call counts vary by trajectory (Appendix~\ref{app:agent}). The surviving corruption comes from two tools whose bank probes miss the change, a stateful tool whose probes ask about a tensor they never created and a batch tool whose schema admits seven distinct probes. Both sit at distance $0.000$ and pass at any budget. An exact equality gate leaves corruption only on these two tools with far fewer probe calls (Figure~\ref{fig:agentview}(d)). Task success is $1.000$ in all four arms and mean judge-rated completion does not fall on the corrupted ones. A metric of apparent task completion therefore misses these wrong values (Appendix~\ref{app:agent}).

\paragraph{Adaptive adversaries and probe secrecy.} Against a committed coverage squatter that returns the gold on a fraction $\rho$ of inputs, a fixed probe set is fooled at every budget while secret subsampling from the bank detects it toward $1-\rho^m$, so detection depends on probe secrecy rather than probe count. Over $88$ cells and $14{,}080$ trials, including white-box squatters written by a code model told how ProbeID works, higher coverage converts identification into abstention and no squatter is selected, as a deterministic gold at distance zero guarantees (Figure~\ref{fig:main}(f), Appendix~\ref{app:adversaries}).

\section{Limitations}
Verification needs a trusted admission reference and probes that reach the difference. A compromised reference leads ProbeID to accept compromised behavior, and drift outside the probe distribution passes unseen, which leaves $9$ of $16$ gated agent tasks tainted. Side effects that never reach a response go unobserved, so behavioral identity is not endpoint identity. The agent gate reads its whole bank at every check, so an adversary that observes the probes and adapts is not covered. A full-bank check costs up to $512$ probe calls and minutes on a slow remote tool, and a refused legitimate update requires a new reference. The held-out replacements transform genuine answers and are not independent attackers. Stateful tools fall outside the independent-probe theorem, small margins in large outputs raise the budget (Appendix~\ref{app:dilution}), and production scale is untested.

\section{Conclusion}
ProbeID tests whether a candidate's canonicalized responses agree with a trusted behavioral reference and abstains when the evidence is inconclusive. Across the reported studies its calibrated gates refuse sufficiently separated replacements and reduce corrupted agent inputs without changing benchmark success, within the limits of reference integrity, probe support and response noise.

\section*{Ethics statement}
ProbeID is a defensive mechanism for agent tool use. It detects when a selected tool's probe-visible responses no longer match its trusted reference, which mitigates rug pull and squatting attacks on open tool registries. A failure mode is misplaced trust in the reference itself. If reference behavior was recorded from an already compromised tool, ProbeID will accept the wrong behavior. Abstention limits silent errors but does not remove this dependency on a clean admission step. Behavioral probing also executes tools, so probes must be confined to read-only or sandboxed calls. The cached studies replay saved responses, the census probes only whitelisted read-only tools, and the stateful and write studies probe reset sandboxes.

\section*{Reproducibility statement}
Every reported number is traceable to a released artifact, with two exceptions stated where they arise. The original judge runs kept no per-case records (Appendix~\ref{app:judge}), and the full raw response corpus will be released upon acceptance (Appendix~\ref{app:impl}). The identity rule and the bounded behavioral distance are stated in Equation~\eqref{eq:dist} and Equation~\eqref{eq:nearest} and given in full, including the canonicalizer and the three response channels, in Appendix~\ref{app:dy}. Theorem~\ref{thm:bound} is stated in the body and Theorem~\ref{thm:abstain} in the appendix, with proofs in Appendices~\ref{app:proof} and~\ref{app:abstain}. The separability condition, the variance-aware refinement and the matching lower bound are in Appendices~\ref{app:sep}, \ref{app:bernstein} and~\ref{app:lower}. Benchmark construction, the squatter families, the probe generator and the seeds are documented in Appendix~\ref{app:impl}. The code and the result files behind every reported number will be released upon acceptance. A README names the result file and the script behind each table and figure, and one script recomputes the main and judge tables, the noise factorial and the agent replay counts from those files. The in-the-wild census of Appendix~\ref{app:census} pins each server to an exact version and records every probe argument and canonicalized response, so the drift replay re-issues identical probes and is checkable against the released per-replay summaries. The agent study of Appendix~\ref{app:agent} releases every recorded tool call with its arguments and the response the agent received, from which its ground truth is computed. It also releases the replay verdicts for all twelve runs, three of each arm, and the gate's own verdict logs, so the corrupted-call counts, the spread across repeats, and the probe cost can be recomputed. The random-probe validation, the sequential gate, the held-out study, the sharper gates and the judge information control are recomputed from the released per-entry distances, the recorded held-out banks and the cached benchmark, each by one script. The held-out study's replacement policies and probe pools are released with it.

\section*{AI use statement}
Generative AI tools were used to assist with writing code, with revising and polishing the writing of this paper, and with generating the icons in Figure~\ref{fig:teaser}. The authors reviewed all AI-assisted work and take full responsibility for the content of this paper.

\bibliography{references}
\bibliographystyle{iclr2027_conference}

\clearpage
\appendix
\let\addcontentsline\probeidaddcontentsline
\etocdepthtag{appendix}
\renewcommand{\thesection}{\AlphAlph{\value{section}}}
\counterwithin{figure}{section}
\counterwithin{table}{section}
\counterwithin{equation}{section}
\section*{Appendix}
Appendix~\ref{app:relwork} extends the related work, and Appendices~\ref{app:proof} to~\ref{app:budgetbounds} prove the results of Section~\ref{sec:theory} and define the distance. Appendices~\ref{app:adversaries} and~\ref{app:judge} support the adversary and judge comparisons of Results~1 and~2, and Appendices~\ref{app:routing} to~\ref{app:cached} the routing, noise and budget studies of Results~1 to~3. Appendices~\ref{app:live} to~\ref{app:dilution} report the live, stateful and large-output studies, Appendices~\ref{app:census} to~\ref{app:banksample} the census, the agent study and the random-probe validation, Appendix~\ref{app:heldout} the prospective held-out study and Appendix~\ref{app:exact} the exact and finite-population gates. Appendix~\ref{app:impl} gives implementation and reproduction details.
\subsection*{Contents}
\begingroup
\footnotesize
\etocsetnexttocdepth{section}
\etocsettagdepth{main}{none}
\etocsettagdepth{appendix}{section}
\etocobeydepthtags
\etocsettocstyle{\raggedright\begin{multicols}{2}}{\end{multicols}}
\etocsetstyle{section}{}{\noindent\hangindent=2.5em\hangafter=1}{\makebox[2.5em][l]{\etocnumber}\etocthename\par}{}
\tableofcontents
\endgroup
\suppressfloats[t]

\section{Behavioral identity in prior work}
\label{app:relwork}
Judging an artifact by what it does or contains, rather than by its declared metadata, is a recurring idea. Conformance testing \citep{chow1978fsm}, query learning \citep{angluin1987lstar}, stateful API fuzzing \citep{atlidakis2019restler}, and network service scanning \citep{lyon2009vscan} all examine implementations through probe responses. Software birthmarks identify programs by intrinsic characteristics they cannot easily fake \citep{tamada2004birthmark,myles2005kgram}, input-output examples retrieve reusable code by sampling behavior \citep{podgurski1993behavior}, and language models are fingerprinted from their responses \citep{pasquini2025llmmap,shao2026zeroprint}. Closest in spirit, Model Equality Testing uses a two-sample test over sampled outputs to ask whether a black-box API still serves a trusted reference model \citep{gao2025modelequality}. Related audits detect silent endpoint substitution \citep{zhu2026rank} or verify which computation produced an output under nondeterminism \citep{karvonen2025difr}, and spoofing attacks show that passively collected fingerprints can be gamed \citep{zhang2026spoofing}. Some methods already combine selection with rejection. LLMmap's closed-set variant attributes a queried endpoint among many known model templates, and its open-set variant thresholds relative distances. In this line of work, an adversary, if present, attacks the verifier. Fingerprint spoofing adapts a model until an auditor mistakes it for another \citep{zhang2026spoofing}, and no guarantee is attached to a decision among candidates authored to be mistaken for a trusted one. ProbeID's setting is $k$ candidates that are schema compatible and description identical by construction, under an explicit drift model, with a margin condition, a probe-budget bound with a matching lower bound on its leading term, and an abstention rule for behaviorally ambiguous tools. It thus differs from model version attribution \citep{pasquini2025llmmap} and program theft detection \citep{tamada2004birthmark} in the adversarial construction of the candidate set and the joint error bound on the decision, not in the act of selecting.

The verification mechanism itself follows established software testing practice. Differential testing runs independent implementations on the same inputs and flags disagreements \citep{mckeeman1998differential}, and Diffy compares a candidate service with two trusted instances, using their disagreement to cancel noise much as ProbeID's self distance does \citep{twitter2015diffy}. Registry-side testing of web services at registration and at discovery, with test scripts attached to the registered service, anticipates the admission reference and the use-time check \citep{tsai2003uddi,bertolino2005audition}. ProbeID adds a calibrated decision and abstention rule to this practice and studies it against description-copying squatters on MCP registries.

Three concurrent defenses for MCP servers observe behavior at run time. MCPShield generates invocations from a tool's metadata and checks the observed behavior against that declaration, including drift over time \citep{zhou2026mcpshield}. TrustShiftProbe audits a single server's responses against a baseline learned while it was trusted, without an oracle \citep{rostamzadeh2026trustshift}. FlowGuard sends schema-valid probes to gather evidence of vulnerabilities, and stops when a round adds no actionable signal \citep{flowguard2607}. Each compares a tool with its own declaration or history, which for TrustShiftProbe is a baseline recorded while the tool was trusted. None probes description-identical candidates on shared inputs, compares their responses with a recorded trusted reference, or bounds the error of selecting among them. The approaches are complementary, since a tool can match its own history and still differ from the reference that was admitted.

\section{Proof of Theorem~\ref{thm:bound}}
\label{app:proof}
Fix the trusted reference $f_{\mathrm{ref}}$ and a candidate $t$. The empirical distance $\Disthat(f_t,f_{\mathrm{ref}})=\frac1m\sum_{i=1}^m Z_i$ averages $m$ terms $Z_i=d_Y(c(f_t(x_i)),c(f_{\mathrm{ref}}(x_i)))\in[0,1]$ that are independent across the i.i.d. probes $x_i\sim D$, with mean $\Dist(f_t,f_{\mathrm{ref}})$. By Hoeffding's inequality, for any $\epsilon>0$,
\[
\Pr\!\big[\,|\Disthat(f_t,f_{\mathrm{ref}})-\Dist(f_t,f_{\mathrm{ref}})|\ge\epsilon\,\big]\le 2\exp(-2m\epsilon^2).
\]
The deployed protocol draws its $m$ probes as a secret subsample without replacement from an attested bank of size $B$ rather than i.i.d. from $D$. Hoeffding's inequality also holds for sampling without replacement from a finite population \citep{hoeffding1963probability}, and \citet{serfling1974} sharpens it by the finite-population correction $1-(m-1)/B$. For the bank, the statement holds with $D$ read as the uniform distribution over the $B$ recorded probes and the margin measured on those probes. A margin on a population distribution transfers to the bank only if the bank was itself drawn i.i.d. from it, at the cost of a second Hoeffding radius over $B$. The guarantee still does not cover the choice of probe distribution. A deterministic enumeration concentrates its support on a predictable set, which the method section excludes and Appendix~\ref{app:adaptive} quantifies through the coverage $\rho$. Apply a union bound over the at most $n+k$ candidates and set the total failure probability to $\delta$, so $2(n+k)\exp(-2m\epsilon^2)\le\delta$, giving $\epsilon=\sqrt{\log(2(n+k)/\delta)/(2m)}$. On the complementary event, which holds with probability at least $1-\delta$, every empirical distance is within $\epsilon$ of its expectation.

Under the margin condition the gold has expected distance $\mu_{t^\star}$ and every wrong candidate has expected distance at least $\mu_{t^\star}+\gamma$. If $\epsilon<\gamma/2$, then the gold empirical distance is below the midpoint $\mu_{t^\star}+\gamma/2$ and every wrong empirical distance is above it, so the gold is strictly nearest and Equation~\eqref{eq:nearest} returns a behaviorally correct tool. The condition $\epsilon<\gamma/2$ is equivalent to $m>\frac{2}{\gamma^2}\log\frac{2(n+k)}{\delta}$, which proves the claim. The dependence on $n+k$ is logarithmic, so $k$ squatters enter the budget only through this factor. $\qquad\blacksquare$

The argument is the finite-class uniform convergence bound of agnostic PAC learning \citep[Corollary~4.6]{shalevshwartz2014understanding} applied to the $n+k$ candidates, and up to constants it is the budget of uniform-sampling PAC best-arm identification \citep{evendar2002pac}. The margin $\gamma$ plays the role of an indifference zone \citep{bechhofer1954single}. We claim no new concentration result, only its application to behavioral verification.

\section{The abstaining algorithm}
\label{app:abstain}
Theorem~\ref{thm:bound} certifies the forced nearest-fingerprint choice. We certify the deployed abstaining rule, Algorithm~\ref{alg:probeid}, and relate $\gamma_{\min}$ to the same confidence radius.

\begin{theorem}[Calibrated abstention]
\label{thm:abstain}
Assume the setting of Theorem~\ref{thm:bound} with margin $\gamma$, the gold present, and at least two candidates. Write $\epsilon=\sqrt{\log(2(n+k)/\delta)/(2m)}$ and run Algorithm~\ref{alg:probeid} with $\gamma_{\min}=2\epsilon$ and $\tau_{\mathrm{hi}}\ge\mu_{t^\star}+\epsilon$. With probability at least $1-\delta$, the algorithm \emph{(i)} does not return a behaviorally wrong candidate and \emph{(ii)} returns a behaviorally correct tool whenever $\gamma\ge 4\epsilon$ and no behavioral twin of the gold is among the candidates. Otherwise, it returns a correct candidate or \bot{} on the same event. This bounds the joint event of a wrong decision and a decision. It does not bound the error rate conditional on deciding by $\delta$.
\end{theorem}

In terms of probabilities, $\Pr(\text{wrong and decide})\le\delta$ under the stated assumptions, and when $\Pr(\text{decide})>0$ the conditional error satisfies only $\Pr(\text{wrong}\mid\text{decide})\le\delta/\Pr(\text{decide})$. An observed error rate among decisions is not a $\delta$-level certificate, and a rule that rarely decides can have a conditional error well above $\delta$.

Let $E$ be the event that $|d_t-\mu_t|\le\epsilon$ for every candidate $t$, where $d_t=\Disthat(f_t,f_{\mathrm{ref}})$ and $\mu_t=\Dist(f_t,f_{\mathrm{ref}})$. The Hoeffding union bound in the proof above gives $\Pr[E]\ge 1-\delta$ for this $\epsilon$. Argue on $E$.

\emph{(i)} Suppose Algorithm~\ref{alg:probeid} returns a candidate, so the nearest passes both gates, $d_{(1)}\le\tau_{\mathrm{hi}}$ and $d_{(2)}-d_{(1)}\ge\gamma_{\min}=2\epsilon$. If the returned candidate were a behaviorally wrong $w$, then $d_{(1)}=d_w$, while the gold has $d_{t^\star}\ge d_{(2)}$ since the gold distance is at least the second smallest once $w$ is smallest. On $E$, $\mu_w\le d_{(1)}+\epsilon$ and $\mu_{t^\star}\ge d_{(2)}-\epsilon$, so $\mu_w-\mu_{t^\star}\le (d_{(1)}-d_{(2)})+2\epsilon\le 2\epsilon-\gamma_{\min}=0$, contradicting $\mu_w-\mu_{t^\star}\ge\gamma>0$. Hence the returned candidate is behaviorally correct, either the gold or a behavioral twin of it.

\emph{(ii)} If $\gamma\ge 4\epsilon$, then on $E$ the gold satisfies $d_{t^\star}\le\mu_{t^\star}+\epsilon\le\tau_{\mathrm{hi}}$, and every wrong $w$ satisfies $d_w\ge\mu_{t^\star}+\gamma-\epsilon$. With no behavioral twin present the next nearest candidate is wrong, so the gold is strictly nearest and the gap is at least $(\mu_{t^\star}+\gamma-\epsilon)-(\mu_{t^\star}+\epsilon)=\gamma-2\epsilon\ge 2\epsilon=\gamma_{\min}$. Both gates pass, so ProbeID returns the gold.

\emph{(iii)} In the remaining regime the gap gate may pass, returning a behaviorally correct tool by \emph{(i)}, or fail and return \bot{}. In neither case is a wrong candidate returned. $\qquad\blacksquare$

The gap gate is the stopping condition of action elimination and of LUCB-style best-arm algorithms \citep{evendar2006action,kalyanakrishnan2012pac}, and abstention is Chow's reject option \citep{chow1970reject}. Selective classification calibrates the conditional risk itself \citep{elyaniv2010selective,geifman2017selective}, whereas Theorem~\ref{thm:abstain} bounds it only by the coverage-dependent $\delta/\Pr(\text{decide})$ above.

The trichotomy needs the gold among the candidates and the stated sampling, margin and threshold conditions. On its probability $1-\delta$ event a returned candidate is correct, the gold is returned once its margin exceeds $4\epsilon$ with no behavioral twin present, and below that margin the rule returns a correct candidate or abstains. The stateful study is empirical evidence outside the independent-probe theorem. Since $\epsilon$ shrinks as $1/\sqrt{m}$, a larger budget lowers the margin sufficient for acceptance at the same error bound.

\paragraph{A single registered candidate.} A deployed gate often sees one server under a trusted name, and that server may be a squatter that replaced the gold, as in the agent study of Appendix~\ref{app:agent}. With one candidate there is no second distance, and Algorithm~\ref{alg:probeid} reduces to its absolute gate, accepting when $d_{(1)}\le\tau_{\mathrm{hi}}$. On the event $E$ for that one candidate, with $\tau_{\mathrm{hi}}=\mu_{t^\star}+\epsilon$, the gold is accepted, and a candidate $w$ is refused whenever $\mu_w-\mu_{t^\star}>2\epsilon$, since then $d_w\ge\mu_w-\epsilon>\tau_{\mathrm{hi}}$. Below that margin the bound gives no protection. A squatter within $2\epsilon$ of the reference may be accepted, and one at distance zero on $D$ always is. The agent gate draws each tool's whole bank, at most $512$ probes, and sets $\tau_{\mathrm{hi}}$ to the tool's self distance plus $\epsilon=\sqrt{\log(2/\delta)/(2m)}$ at $\delta{=}0.05$. The threshold is $0.060$ for a full bank at zero self distance and $0.51$ for a tool whose schema admits only seven distinct probes. The two tools that survive the gate sit at distance $0.000$ and are accepted, as this bound allows (Appendix~\ref{app:agent}).

\paragraph{A sequential gate.} A fixed budget spends the whole bank on every check, even when a few probes already settle it. The sequential gate instead reads nested prefixes of one secret random permutation of the bank at checkpoints $m_1<\dots<m_J$ and splits the error budget evenly, $\delta_j=\delta/J$ and $\epsilon_j=\sqrt{\log(2/\delta_j)/(2m_j)}$. With $d^{(j)}$ the mean distance over the first $m_j$ probes, it refuses at the first checkpoint where $d^{(j)}>s+\epsilon_j$, accepts at the first where $d^{(j)}+\epsilon_j\le s+\gamma_{\mathrm{res}}$, and reports the check as inconclusive if the last checkpoint does neither. The resolution $\gamma_{\mathrm{res}}$ is an indifference zone chosen by the deployer.

\begin{proposition}[Sequential gate]
\label{prop:sequential}
Let $\mu$ be a candidate's expected distance on the bank distribution and $s$ the admission self distance. With probability at least $1-\delta$, the sequential gate does not refuse the candidate when $\mu\le s$ and does not accept it when $\mu>s+\gamma_{\mathrm{res}}$. On every draw it decides no later than the first checkpoint with $2\epsilon_j\le\gamma_{\mathrm{res}}$, if one exists.
\end{proposition}

Let $E$ be the event that $|d^{(j)}-\mu|\le\epsilon_j$ at every checkpoint. Each prefix is a uniform sample without replacement from the bank, so Hoeffding's inequality bounds each failure by $\delta_j$, and a union bound over the $J$ checkpoints gives $\Pr[E]\ge1-\delta$. On $E$, if $\mu\le s$ then $d^{(j)}\le s+\epsilon_j$ at every checkpoint and the gate never refuses, and if $\mu>s+\gamma_{\mathrm{res}}$ then $d^{(j)}+\epsilon_j\ge\mu>s+\gamma_{\mathrm{res}}$ and it never accepts. At a checkpoint with $2\epsilon_j\le\gamma_{\mathrm{res}}$, a gate that does not refuse has $d^{(j)}\le s+\epsilon_j$, hence $d^{(j)}+\epsilon_j\le s+\gamma_{\mathrm{res}}$, and it accepts. $\qquad\blacksquare$

A candidate with $s<\mu\le s+\gamma_{\mathrm{res}}$ may be accepted, so the resolution plays the role that $2\epsilon$ plays at a fixed budget, now fixed in advance rather than set by the budget. The same splitting makes Algorithm~\ref{alg:probeid} sequential. With $\epsilon_j=\sqrt{\log(4/\delta_j)/(2m_j)}$ for two candidates and a decision at the first checkpoint whose two gates pass, part \emph{(i)} of Theorem~\ref{thm:abstain} holds at every checkpoint on the same kind of event, so the joint probability of a wrong decision stays at most $\delta$. The even split is the simplest valid choice. Time-uniform confidence sequences \citep{howard2021timeuniform} and concentration bounds for sampling without replacement \citep{bardenet2015concentration} give tighter radii, the latter shrinking to zero as $m$ approaches the bank size. On the agent banks with checkpoints $8,16,\dots,512$ and $\delta{=}0.05$, the sequential equality gate refuses every reachable replacement after $12.5$ probes on average, where the fixed gate needs $64$ to stop accepting them, and at $\gamma_{\mathrm{res}}{=}0.2$ accepts $76$ of the $77$ golds within $128$ probes. It refuses none of the golds and leaves convert\_batch, whose bank holds seven probes, inconclusive. No observed rate breaks the proposition's bounds on these banks or in the held-out study of Appendix~\ref{app:heldout}.

\section{Sources of the behavioral margin}
\label{app:sep}
Theorem~\ref{thm:bound} takes the margin $\gamma$ as given. The margin depends on the probe distribution and on each candidate's responses. This section gives conditions for a positive margin and a way to measure it for tested candidates.

For a deterministic gold whose canonical response matches the reference for $D$-almost every probe, $\mu_{t^\star}=0$. Fix a wrong candidate $w$ and define $\Delta_w=\{x:d_Y(c(f_w(x)),c(f_{\mathrm{ref}}(x)))>0\}$ with mass $p_w=\Pr_{x\sim D}[x\in\Delta_w]$. The distance is zero on $\Delta_w^c$ and positive on $\Delta_w$. Hence $\mu_w=\int_{\Delta_w} d_Y\,dD\le p_w$, and $\mu_w>0$ if and only if $p_w>0$. This distance-based definition matters because distinct canonical values can have zero implemented distance.

\begin{proposition}[Separability]
\label{prop:sep}
With a deterministic pointwise faithful gold and a finite candidate pool, $\gamma=\min_w\mu_w>0$ if and only if every squatter has positive distance from the reference on a set of positive probe mass, $\min_w p_w>0$. A candidate with $p_w=0$ is indistinguishable under $d_Y$ and $D$. When the gold is also present, the margin gate abstains on their tied observations.
\end{proposition}

The characterization is qualitative by necessity. If the canonical responses reachable on $D$ take finitely many structural distance values, for instance bounded-size outputs, or a scalar or binary channel where the smallest positive distance is $d_{\min}=1$, the bound sharpens to the quantitative $\mu_w\ge d_{\min}\,p_w$ with $d_{\min}$ the smallest positive value $d_Y$ attains. On variable-size outputs no uniform $d_{\min}$ exists, and the margin can be small even at large divergence mass. Appendix~\ref{app:dilution} measures this dilution on a live server. A format-faithful squatter that perturbs $f$ of $N$ response fields leaves a per-probe margin $\gamma=f/(12N)$, so no schema-level constant bounds $\gamma$ away from zero, and in that regime the abstention rule, not the identity rule, carries the guarantee.

Three consequences follow. First, probe design controls the margin. Boundary and enumeration values seek differences, but schema validity does not ensure positive $D$-mass on every $\Delta_w$. The agent study exhibits missing state prerequisites and a nested schema whose probes miss the altered behavior (Appendix~\ref{app:agent}). Second, if a squatter has zero distance on a fraction $\rho=\Pr_D[x\in\Delta_w^c]$ of inputs, then $p_w=1-\rho$. The margin is at least $d_{\min}(1-\rho)$ when a positive minimum distance exists. Third, zero margin means that at least one candidate is indistinguishable under this distance and probe distribution. It does not imply identical behavior outside the probe support or in fields the comparison ignores.

\section{The behavioral distance \texorpdfstring{$d_Y$}{dY}}
\label{app:dy}
The canonicalizer $c$ drops, at any depth and regardless of case, twelve volatile keys, \texttt{timestamp}, \texttt{request\_id}, \texttt{requestid}, \texttt{id}, \texttt{trace\_id}, \texttt{traceid}, \texttt{cursor}, \texttt{created\_at}, \texttt{updated\_at}, \texttt{nonce}, \texttt{etag} and \texttt{date}. It sorts the remaining keys, rounds floats to six decimals, keeps lists in order and booleans as they are, and applies no whitespace or Unicode normalization. A response delivered as JSON text is parsed first. Every response carries a status, set to success when the server reports none, and a state delta, empty when none is observed. On canonicalized values the distance reduces to a recursive structural distance $\delta\in[0,1]$. For two scalars $\delta(a,b)=\mathbb{1}[a\neq b]$, an integer and a float of equal value count as equal, and a type mismatch gives $\delta=1$. For maps over the key union $K=\mathrm{keys}(a)\cup\mathrm{keys}(b)$,
\[
\delta(a,b)=\frac{1}{|K|}\sum_{\kappa\in K}
\begin{cases}\delta(a_\kappa,b_\kappa) & \kappa\in a\cap b,\\ 1 & \text{otherwise,}\end{cases}
\]
and for lists with $L=\max(|a|,|b|)$, $\delta(a,b)=\frac{1}{L}\sum_{i<L}\delta(a_i,b_i)$ where a missing index is charged $1$. By convention $\delta(a,b)=0$ when $K=\emptyset$ (two empty maps) or $L=0$ (two empty lists), so an empty container matches itself at distance $0$ and $\delta$ is total on canonicalized values. A canonicalized response carries three observable channels (output, status and state delta), combined by a probabilistic union so that any channel difference registers,
\[
d_Y(a,b)=1-\prod_{c\,\in\,\{\text{out},\,\text{stat},\,\Delta\}}\big(1-\delta(a_c,b_c)\big).
\]
Each $d_Y\in[0,1]$ is symmetric and is $0$ for identical responses, and the empirical behavioral distance is the probe mean $\Disthat(f,g)=\frac{1}{m}\sum_{i=1}^m d_Y(c(f(x_i)),c(g(x_i)))$, so the $[0,1]$ range assumed by Theorem~\ref{thm:bound} holds. The probabilistic union keeps a genuine output difference from being diluted when status and state delta are constant, the common case for read-style tools. Without a canonicalizer, a simpler $0/1$ equality or hash comparison of these responses collapses under volatile fields while the bounded $d_Y$ remains effective (Appendix~\ref{app:baselines}).

\section{Variance-aware budget and a matching lower bound}
\label{app:budgetbounds}
\subsection{Variance-aware budget}
\label{app:bernstein}
Theorem~\ref{thm:bound} bounds the worst case over $d_Y\in[0,1]$, which is loose when the per-probe terms $Z_i=d_Y(c(f_t(x_i)),c(f_{\mathrm{ref}}(x_i)))$ have small variance. Suppose every candidate, including a noisy gold with mean $\mu_{t^\star}$, has per-probe variance $\mathrm{Var}(Z_i)\le\sigma^2$. By Bernstein's inequality, $\Pr[|\Disthat-\Dist|\ge\epsilon]\le 2\exp\!\big(-\tfrac{m\epsilon^2}{2\sigma^2+2\epsilon/3}\big)$. Repeating the union bound over $n+k$ candidates and the midpoint margin step $\epsilon<\gamma/2$ gives the budget
\[
m \;>\; \frac{8\sigma^2}{\gamma^2}\log\frac{2(n+k)}{\delta} \;+\; \frac{4}{3\gamma}\log\frac{2(n+k)}{\delta},
\]
whose leading term is $O(\sigma^2\gamma^{-2}\log((n+k)/\delta))$ and recovers the noiseless rate as $\sigma\to0$ up to the additive $1/\gamma$ term. The variance admits a plug-in estimate from the probe sample, $\hat\sigma^2=\frac1m\sum_{i=1}^m (Z_i-\bar Z)^2$, which gives a run-time diagnostic, and a high-confidence certificate would add the standard empirical Bernstein variance term. Result~2 sweeps $\sigma$ through the controlled channel and reports $m^\star\gamma^2$ growing with $\sigma^2$ as this predicts.

\subsection{A lower bound matching the leading term}
\label{app:lower}
Theorem~\ref{thm:bound} and Appendix~\ref{app:bernstein} give an upper bound of order $\sigma^2/\gamma^2$ on the probe budget. Over independent candidate channels this rate is minimax optimal. No identity rule separates a gold tool from a squatter at margin $\gamma$ with fewer than order $\sigma^2/\gamma^2$ probes, so the leading term of the budget cannot be reduced beyond constants. The construction below makes the candidate channels independent, the regime in which the lower bound is tight. Correlated drift across squatters falls outside this independent construction, so the minimax floor does not formally cover it. Result~3 observes live that such correlation only makes the gold easier to single out, but that is an empirical observation and does not follow from the lower bound.

Fix a single candidate ($k{=}1$) and reduce identity to a two-point test. Each probe $x_i\sim D$ yields the mismatch indicator $Z_i=d_Y(c(f(x_i)),c(f_{\mathrm{ref}}(x_i)))\in\{0,1\}$. Under the gold hypothesis $H_0$ the candidate is the reference up to noise, so $Z_i\sim\mathrm{Bernoulli}(p)$ for some $p\in(0,1)$, with self distance mean $p$ and per-probe variance $\sigma^2=p(1-p)$. Under the squatter hypothesis $H_1$ the candidate diverges on an extra $\gamma$ fraction of probes, so $Z_i\sim\mathrm{Bernoulli}(p+\gamma)$ with $p+\gamma\le 1$, a behavioral distance $\gamma$ to the reference. An identity rule with budget $m$ is a test $\psi$ on the i.i.d. sample $Z_1,\dots,Z_m$ deciding $H_0$ against $H_1$. The binary mismatch indicator is a realizable instance of the bounded behavioral distance. A tool whose canonical response on each probe is either equal to the reference or maximally distant makes $d_Y\in\{0,1\}$, which scalar-output tools such as get-sum already satisfy. A minimax lower bound proved on an instance inside the model class is a valid lower bound for the class, so the rate below cannot be beaten in general even though $d_Y$ may take fractional values on other tools.

By Le Cam's two-point lemma \citep{tsybakov2009nonparametric}, every such test has average error
\[
\tfrac12\big(\Pr\nolimits_{H_0}[\psi{=}1]+\Pr\nolimits_{H_1}[\psi{=}0]\big)\ \ge\ \tfrac12\big(1-\mathrm{TV}(P_0^{\otimes m},P_1^{\otimes m})\big),
\]
where $P_0=\mathrm{Bernoulli}(p)$ and $P_1=\mathrm{Bernoulli}(p+\gamma)$. Pinsker's inequality with tensorization gives $\mathrm{TV}(P_0^{\otimes m},P_1^{\otimes m})\le\sqrt{\tfrac12\,m\,\mathrm{KL}(P_0\Vert P_1)}$, and a second-order expansion of the Bernoulli divergence, whose first-order term vanishes at $\gamma=0$, yields $\mathrm{KL}(P_0\Vert P_1)=\frac{\gamma^2}{2p(1-p)}+O(\gamma^3)=\frac{\gamma^2}{2\sigma^2}(1+o(1))$. Combining the two bounds,
\[
\mathrm{TV}(P_0^{\otimes m},P_1^{\otimes m})\ \le\ \frac{\gamma}{2\sigma}\sqrt{m}\,(1+o(1)).
\]
For the average error to fall to $\delta$ the total variation must reach $1-2\delta$, which forces
\[
m\ \ge\ \frac{4(1-2\delta)^2\,\sigma^2}{\gamma^2}\,(1+o(1)).
\]
Hence $m=\Omega(\sigma^2/\gamma^2)$ probes are necessary, matching the Bernstein upper bound of Appendix~\ref{app:bernstein} in the $\sigma^2/\gamma^2$ factor.

The $\log(n+k)$ factor in Theorem~\ref{thm:bound} is also necessary, by a multiple-hypothesis argument. Take $M=n+k$ candidates and $M$ registry configurations, where configuration $j$ makes candidate $j$ the gold, with per-probe mismatch law $\mathrm{Ber}(p)$ against the reference, and every other candidate a squatter at margin $\gamma$, with law $\mathrm{Ber}(p+\gamma)$. A verifier that draws $m$ shared probes and observes all candidates must name the gold, an $M$-ary test. Between configurations $j$ and $j'$ only candidates $j$ and $j'$ change law, so with the symmetrized divergence $J(a,b)=\mathrm{KL}(\mathrm{Ber}(a)\Vert\mathrm{Ber}(b))+\mathrm{KL}(\mathrm{Ber}(b)\Vert\mathrm{Ber}(a))$ the observation laws satisfy
\[
\mathrm{KL}(P_j\Vert P_{j'})=m\,J(p,p{+}\gamma)=\frac{m\gamma^2}{\sigma^2}(1+o(1)).
\]
By Fano's inequality with the mutual information bounded by the largest pairwise divergence, any rule has error
\[
\Pr[\hat\jmath\neq j]\ \ge\ 1-\frac{m\gamma^2/\sigma^2+\log 2}{\log M},
\]
so identifying the gold with constant probability requires $m=\Omega\!\big(\frac{\sigma^2}{\gamma^2}\log(n+k)\big)$. The variance-aware budget of Appendix~\ref{app:bernstein} is thus tight in its leading $\sigma^2\gamma^{-2}\log(n+k)$ term, and Theorem~\ref{thm:bound}, whose worst-case rate takes $\sigma^2$ of constant order, is tight in $\gamma$ and the candidate count. The lower bound does not capture the additive $O(\gamma^{-1}\log(n+k))$ Bernstein term, so the match is in the leading term, not the full finite-sample rate. The two constructions serve different roles. The $k{=}1$ two-point test isolates $\sigma^2/\gamma^2$, while the $M$-ary independent-channel Fano packing yields the product with $\log(n+k)$. Both the $\sigma^2/\gamma^2$ dependence and the $\log(n+k)$ factor are minimax floors over the independent-channel class, attained by the constructions above, so no identity rule removes them uniformly. They do not imply that every registry is this hard, and probe secrecy is no substitute for probes that reach the difference. \citet{kaufmann2016complexity} develop instance-dependent lower bounds of this kind for best-arm identification.

\section{Adaptive and independently authored squatters}
\label{app:adversaries}
\subsection{Randomized probes against committed-coverage adversaries}
\label{app:adaptive}
The bound above assumes the probes are i.i.d. from a distribution $D$, not chosen after seeing the verifier. We now allow an adaptive squatter that returns the gold response on a committed region $C$ of inputs and drifts elsewhere, where $C$ covers a fraction $\rho=\Pr_{x\sim D}[x\in C]$ of the probe distribution. This committed-coverage adversary may know the probe distribution and the verifier, but it commits $C$ before the probes are drawn. An adversary that observes the drawn probes before responding is outside the model. If the verifier draws $m$ probes i.i.d. from $D$, the squatter evades only when every probe lands in $C$, which happens with probability $\rho^m$, so ProbeID detects it with probability at least $1-\rho^m$, and $m\ge\log(1/\delta)/\log(1/\rho)$ probes give detection probability $1-\delta$. A fixed probe set that the adversary has already seen is the case $\rho=1$ on that set and is evaded with certainty. Fresh unpredictable probes give exponential decay in $m$ only when the committed region has $\rho<1$, and as $C$ covers almost all of $D$ the squatter approaches a twin on that distribution, although differences outside its support can remain harmful. Against an adaptive attacker the probes must be secret as well as numerous, and the geometric bound makes this precise within a specific limit. Since $\rho$ is the adversary's coverage of the realized probe distribution, the guarantee requires that $D$ have support the adversary cannot enumerate and cover, and that $C$ be committed without observing the drawn probes. A squatter that reverse-engineers a predictable, small-support probe generator, for instance a verifier that only probes the schema boundary values enumerated by M1, can set $C$ to that support and drive $\rho\to1$, making the bound vacuous. The defense depends on drawing from a large, high-entropy input space, as the live studies do, and not only on the probe count. The probe generator ablation of Appendix~\ref{app:probegen} confirms both halves of this caveat on a live server. Fresh schema-valid probes recover identity, and the M1 boundary set over-covers a fixed drift region, as the small-support concern predicts.

\paragraph{The attested bank on a live server.} Fresh probes must be available without live access to the trusted tool, which the threat model excludes. We rerun the live committed-coverage study with the deployed protocol. At admission a bank of $B{=}256$ schema-valid probes and their canonicalized gold responses is recorded. At selection each trial secretly subsamples $m$ bank probes, the reference fingerprint is read from the recorded bank entries, and only the current candidates are probed live. Over $160$ trials per cell, detection matches the fresh-probe results. At $\rho{=}0.5$, hit@1 rises $0.700$, $0.956$, $1.000$, $1.000$ as $m$ goes $2,4,8,16$ (the live fresh-probe run measured $0.95$ at $m{=}8$), and at $\rho{=}0.9$ it rises $0.163$, $0.362$, $0.619$, $0.844$, tracking $1-\hat\rho^{\,m}$ at the bank's realized coverage $\hat\rho\approx0.887$ of the nominal $0.9$. The fixed-probe control stays at $0.00$ at every budget, so detection depends on the secrecy of the subsample, not on bank size. An attacker who commits its mimicry region before the bank is recorded gains nothing from the bank being finite, and the trusted tool is queried only during admission. Repeated exposure of the same bank can still reveal it (Appendix~\ref{app:agent}).

\paragraph{Coverage and escalation.} Two coverage levels alone do not establish whether a squatter that covers enough of the input space is eventually selected or only escalated. We sweep $\rho$ from $0.1$ to $0.99$ against budgets $m\in\{2,4,8,16\}$, $160$ trials per cell, and score the deployed abstaining rule with its three outcomes rather than the forced choice used above. Table~\ref{tab:adaptivesweep} reports the bank protocol.

\begin{table}[t]
\centering
\caption{\textbf{An adaptive squatter induces abstention without certification while the gold is a candidate} (live get-sum tool, bank of $256$ probes, $160$ trials per cell). Each entry is correct / abstain, and the misidentification rate is $0.000$ in every cell. Raising the committed coverage $\rho$ converts correct identifications into escalations, and a larger secret subsample converts them back.}
\label{tab:adaptivesweep}
\small
\setlength{\tabcolsep}{5pt}
\begin{tabular}{lcccc}
\toprule
$\rho$ & $m{=}2$ & $m{=}4$ & $m{=}8$ & $m{=}16$ \\
\midrule
$0.1$  & 1.000 / 0.000 & 1.000 / 0.000 & 1.000 / 0.000 & 1.000 / 0.000 \\
$0.3$  & 0.950 / 0.050 & 0.981 / 0.019 & 1.000 / 0.000 & 1.000 / 0.000 \\
$0.5$  & 0.700 / 0.300 & 0.956 / 0.044 & 1.000 / 0.000 & 1.000 / 0.000 \\
$0.7$  & 0.425 / 0.575 & 0.794 / 0.206 & 0.938 / 0.062 & 0.994 / 0.006 \\
$0.9$  & 0.163 / 0.838 & 0.362 / 0.637 & 0.619 / 0.381 & 0.844 / 0.156 \\
$0.95$ & 0.087 / 0.912 & 0.163 / 0.838 & 0.306 / 0.694 & 0.550 / 0.450 \\
$0.99$ & 0.006 / 0.994 & 0.013 / 0.988 & 0.044 / 0.956 & 0.138 / 0.863 \\
\bottomrule
\end{tabular}
\end{table}

The sweep separates two quantities that a single number hides. Coverage drives escalation. At $\rho{=}0.99$ and $m{=}2$ the squatter forces abstention on $0.994$ of trials, and a larger budget recovers only part of it, to $0.863$ at $m{=}16$, because at that coverage the squatter agrees with the gold almost everywhere and is close to the behavioral twin the abstention rule exists for. Coverage does not lead to the squatter being selected. Across all $88$ cells of the sweep and both probe modes, $14{,}080$ trials in total, the misidentification rate is $0.000$. The fixed-probe control is the extreme case, abstaining on $1.000$ of trials at every $\rho$ and every $m$, again with no misidentification. A fully predictable probe set is evaded completely, and the failure is total escalation, not a wrong selection. The deterministic gold is among the candidates and has zero distance, so the attack changes abstention rather than which candidate is nearest. This experiment does not establish the same outcome for a replaced endpoint without the gold (Theorem~\ref{thm:abstain}).

Two facts limit the interpretation of the zero misidentification rate. The gold is a live deterministic server whose distance to its own recorded bank is $0.00$ in all $14{,}080$ trials, and the selector abstains when the gap falls below $\gamma_{\min}$, so no squatter can be strictly nearer than the gold. Here a wrong certification is excluded by construction, and the sweep measures the split between identification and abstention. Nor does the zero carry over to a squatter that replaces the gold. With no gold among the candidates the decision is the one-sided gate of Appendix~\ref{app:abstain}, and a squatter is accepted whenever every drawn probe falls in its mimicry region, with probability at least $\rho^m$, the evasion term of the bound at the start of this appendix.

\subsection{Independently authored and verifier-aware squatters}
\label{app:llmsquatter}
The benchmark squatters come from two algorithmic families we wrote, output-value drift and silent field drift, the axis the structural distance measures. To go beyond these families, a separate code model generates squatter behaviors under two threat levels. This broadens the tested attacks but does not cover all adversaries. A \emph{blind} attacker sees only the gold tool card, schema, and a sample response, and writes a plausibly drifted reimplementation. A \emph{verifier-aware} attacker is also told how ProbeID works, including canonicalization, the bounded structural distance and the $\tau_{\mathrm{hi}}/\gamma_{\min}$ gates, and is instructed to minimize detectable divergence while still altering behavior. The squatter designs are the model's own and are released. We run the deployed abstaining selector, with unchanged scoring code, on $50$ gold benchmark tools and on four live tools of the everything reference server.

\begin{table}[t]
\centering
\caption{\textbf{No misidentification of an independently authored squatter, including a white-box one, with the gold among the candidates} ($50$ gold benchmark tools). A blind code-model squatter is detected or causes abstention when it is a behavioral twin. A verifier-aware squatter that knows the distance and thresholds can only drive ProbeID to abstain, since a deterministic gold at distance zero cannot be undercut, and the misidentification rate stays $0$ over these cases.}
\label{tab:llmsquatter}
\small
\setlength{\tabcolsep}{4pt}
\begin{tabular}{llccc}
\toprule
attacker & $k$ & hit@1 & abstain & misid \\
\midrule
blind & 1 & 0.86 & 0.14 & 0.00 \\
blind & 8 & 0.84 & 0.16 & 0.00 \\
verifier-aware & 1 & 0.42 & 0.58 & 0.00 \\
verifier-aware & 8 & 0.10 & 0.90 & 0.00 \\
\bottomrule
\end{tabular}
\end{table}

Table~\ref{tab:llmsquatter} reports the benchmark arm. Against the blind attacker ProbeID holds hit@1 at $0.84$ to $0.86$ across $k$, abstaining on the remainder, with a misidentification rate of $0.00$. An independently written squatter is still separated in behavior space, so the result does not depend on the two mutation families. The verifier-aware attacker is the sharper test. Told the distance and the thresholds, it shrinks its canonical divergence and drives hit@1 down to $0.10$ at $k{=}8$, but the loss in hit@1 corresponds to increased abstention, from $0.58$ at $k{=}1$ to $0.90$ at $k{=}8$, and the misidentification rate stays $0.00$ at every $k$. Knowing the verifier lets a squatter approach a behavioral twin and force an escalation, but in these cases it was never selected. The gold here is its own cached reference at distance zero, so no squatter can be strictly nearer, and the designs change only how much identification turns into abstention. The theorem's bound does not apply to these fixed probes. Across the four squatter counts, the $103$ benchmark cases that the verifier-aware model drove to zero canonical divergence are such twins, and ProbeID abstained on all of them. The live arm agrees. A blind squatter on the four live tools is caught at hit@1 $1.00$, and a verifier-aware one falls to $0.17$ at $k{=}8$ while its misidentification rate stays $0.00$.

\section{A behavioral LLM-judge baseline}
\label{app:judge}
The description routers of Table~\ref{tab:main} collapse to chance because every candidate advertises the gold description, which raises the question of whether a baseline that reads behavior instead of text can match the bounded behavioral distance. The LLM judge runs Qwen2.5-7B. One request per case lists the trusted reference responses and every candidate's responses on the same probes ProbeID uses, in the case's shuffled candidate order, each serialized response cut to $200$ characters. The judge answers with one integer, generating at most $8$ tokens at temperature $0$, and names the candidate whose behavior matches the reference, while $0$ or an unparseable answer counts as an abstention. The prompt is released with the code. The judge sees no descriptions, so the byte-identical rug pull does not defeat it and it does not collapse to $1/(k+1)$. It reads the same probes as ProbeID but only the first $200$ characters of each response, while ProbeID compares complete canonicalized responses, so the comparison does not isolate the distance rule. We evaluate on the described pool of Table~\ref{tab:main}, now with 300 gold tools per seed over three seeds.

The 7B judge performs well above chance. It reaches $0.973$ at $k{=}1$ and falls monotonically as candidates accumulate, to $0.844$ at $k{=}8$, while ProbeID stays at $0.952$ for $k\ge2$, a gap that widens from $2.7$ to $10.8$ points. At $k{=}8$ the 7B judge sits at $0.844$ with a $95\%$ exact binomial interval of $[0.819, 0.868]$ against ProbeID at $0.952$ with $[0.936, 0.965]$, over $n{=}900$ case-runs. These intervals are descriptive. They treat case-runs as independent, although a gold can recur across seeds because the golds of each seed are drawn from one pool of $402$ tools, and the original judge runs kept no per-case records from which a tool-level interval could be computed.

\begin{table}[t]
\centering
\caption{\textbf{Behavioral judges against ProbeID on the judge cohort} ($300$ gold per seed, $3$ seeds, $900$ case-runs per $k$, same pool as Table~\ref{tab:main}). The judges read the probe responses, not the descriptions, so they stay well above the $1/(k+1)$ random bound. The judge-view column is an exact-match rule that reads each response as the judges' prompt shows it, cut to $200$ characters. Best non-oracle result in bold.}
\label{tab:judge}
\small
\setlength{\tabcolsep}{5pt}
\begin{tabular}{lcccccc}
\toprule
$k$ & ProbeID & judge view, exact & 7B judge & 32B judge & Random & Oracle \\
\midrule
1 & \textbf{1.000} & 0.997 & 0.973 & 0.996 & 0.498 & 1.000 \\
2 & 0.952 & 0.951 & 0.923 & \textbf{0.968} & 0.352 & 1.000 \\
4 & 0.952 & 0.951 & 0.906 & \textbf{0.967} & 0.190 & 1.000 \\
8 & \textbf{0.952} & 0.951 & 0.844 & 0.950 & 0.103 & 1.000 \\
\bottomrule
\end{tabular}
\end{table}

\paragraph{Information hidden by truncation.} The effect of truncation can be measured without a model. We rebuilt every case of this cohort from the cache and scored an exact-match rule with strict tie-breaking on each response as the judge's prompt renders it. The rebuild reproduces ProbeID's and the random selector's stored hit@1 at every $k$. Between $498$ and $512$ of the $900$ case-runs per $k$ carry a response longer than the cut, yet the judge-view rule reaches $0.997$ at $k{=}1$ and $0.951$ at $k\ge2$ (Table~\ref{tab:judge}), against $1.000$ and $0.952$ for the same rule on complete responses. The cut hides the only distinguishing difference in $3$ case-runs at $k{=}1$ and in $1$ at each larger $k$. On this cohort truncation removes little information needed for exact matching. The 32B judge's $0.950$ at $k{=}8$ is close to that rule's $0.951$ on the same view, and at $k\in\{2,4\}$ the judge exceeds it, since strict tie-breaking scores the rule's ties as misses.

\begin{table}[t]
\centering
\caption{\textbf{The 32B judge on complete responses matches the truncated condition where it can read them.} Same $900$ case-runs per $k$ and one serving session for both conditions. Sent counts the case-runs whose complete-response prompt fits the $32{,}768$-token context. The first four columns are hit@1 on the sent case-runs, and the last counts unsent case-runs as misses.}
\label{tab:judgefull}
\small
\setlength{\tabcolsep}{4.5pt}
\begin{tabular}{@{}lcccccc@{}}
\toprule
$k$ & sent & 200 characters & complete & ProbeID & complete, all \\
\midrule
1 & 827 & 0.995 & 0.998 & 1.000 & 0.917 \\
2 & 817 & 0.965 & 0.961 & 0.947 & 0.872 \\
4 & 792 & 0.962 & 0.957 & 0.946 & 0.842 \\
8 & 746 & 0.946 & 0.945 & 0.946 & 0.783 \\
\bottomrule
\end{tabular}
\end{table}

\paragraph{Complete responses.} We reran the 32B judge on the same $3{,}600$ case-runs with each response left uncut, beside the $200$-character condition in the same serving session, and kept every choice (Table~\ref{tab:judgefull}). The truncated condition reproduces the values of Table~\ref{tab:judge} exactly. A complete-response prompt longer than the judge's native $32{,}768$-token context cannot be sent, which happens in $73$, $83$, $108$ and $154$ of $900$ case-runs at $k{=}1$, $2$, $4$ and $8$, and $20$ to $56$ of them exceed even $128$k tokens. On case-runs whose prompts fit the context window, the judge on complete responses scores within $0.006$ of the truncated condition at every $k$, and every gold-clustered $95\%$ interval of the difference includes zero. Counted as misses, the unsent case-runs lower complete-response hit@1 to $0.783$ at $k{=}8$, while ProbeID reads every case-run and scores at least $0.981$ on the ones the judge cannot send. A decision costs $1.95$ seconds on truncated prompts and $3.76$ seconds on complete ones, at concurrency $16$ on one $46$GB GPU and excluding probe calls. Where the judge can read complete responses they neither help nor hurt it, and its context limit, which ProbeID does not have, excludes the rest.

Scaling the judge closes the accuracy gap. A Qwen2.5-32B judge on the same truncated evidence reaches $0.996$, $0.968$, $0.967$, and $0.950$ hit@1 at $k\in\{1,2,4,8\}$, within a point or two of ProbeID and ahead at $k\in\{2,4\}$, while the random selector and oracle are bit for bit identical to the 7B run, since the pool and gold are unchanged. The study places ProbeID at $0.952$ on this cohort against $0.956$ on the 150-gold headline cohort. A sufficiently large behavioral judge recovers identity, so the router collapse of Table~\ref{tab:main} comes from the copied descriptions and not from the behavioral channel. ProbeID and the 32B judge differ in the cost of the decision step, not in accuracy. Appendix~\ref{app:determinism} measures $3.40$ seconds per decision for the judge on a dedicated $46$GB GPU against $3.527$ milliseconds for the bounded distance on one CPU core, both excluding probe calls. Access to behavior is necessary but not sufficient at 7B, and a large judge or the bounded behavioral distance closes the remaining gap to the oracle.

\subsection{Decision reproducibility and inference cost}
\label{app:determinism}
A deterministic decision would distinguish ProbeID from a model judge only if the judge's decisions varied across identical requests. We replayed the judge's requests to measure this and also report the cost of each decision step.

The judge is queried at temperature $0$, which is nominally greedy. Greedy decoding under a real serving stack is not by itself a reproducibility guarantee, since continuous batching, prefix caching, and reduction order inside fused kernels can all vary with server state, and the temperature pins none of them. We replayed identical requests and recorded the \emph{chosen candidate}, not only whether it was a hit, because two passes that both miss while naming different wrong candidates disagree in a way a hit vector cannot show. ProbeID's choice was recomputed on the same cases in the same loop as a control.

Over $900$ cases at $k{=}8$, three seeds, replayed three times against one server at concurrency $16$, the judge produced zero decision changes, and per-seed hit@1 was identical to four decimals ($0.9400$, $0.9367$, $0.9733$, mean $0.9500$). After a server restart, three further passes over $300$ cases at concurrencies $1$, $16$, and $4$ returned identical choice vectors, again with zero changes. Aggregate hit@1 at $k{=}8$ was also $0.950$ under vLLM $0.19.1$, which produced the $32$B numbers of Appendix~\ref{app:judge}, and $0.9500$ under $0.24.0$ above, although choice vectors were not compared across versions. The deterministic ProbeID control was unchanged. These observations are limited to the configurations we varied and show no reproducibility advantage for ProbeID.

The cost of the decision step was measured on the same machine and the same cases. The judge takes $3.40$ seconds per decision on a dedicated $46$GB GPU at concurrency $16$ ($9186$ seconds for $2700$ decisions). ProbeID takes $3.527$ milliseconds per decision on one CPU core, timing the identity rule alone with case construction excluded, over the same $900$ cases at $k{=}8$ and $m{=}12$. The judge figure also assumes a $46$GB accelerator, a served $32$B checkpoint, and an inference stack whose version becomes part of the verifier's trusted base, while the bounded distance is arithmetic over responses already collected. Both figures exclude the probe calls themselves, which the judge needs as much as the bounded distance does, and on real servers those calls are the larger cost, so neither figure is an end-to-end latency. Over the $1{,}248$ probe responses recorded at $t_0$ on the $16$ census servers of Appendix~\ref{app:census}, the median call returns in $1.3$ milliseconds, but $15.5\%$ take longer than a second, the $90$th percentile is $1.8$ seconds, and four of the $16$ servers have a median above one second. Wherever a tool calls a slow remote service, the end-to-end cost of a decision is set by executing $m$ probes on each candidate, and the gap in the decision step does not carry over to it.

\section{Description routing and natural cases}
\label{app:routing}
\subsection{Beyond identical descriptions}
\label{app:descsim}
\begin{figure}[t]
\centering
\includegraphics[width=\textwidth]{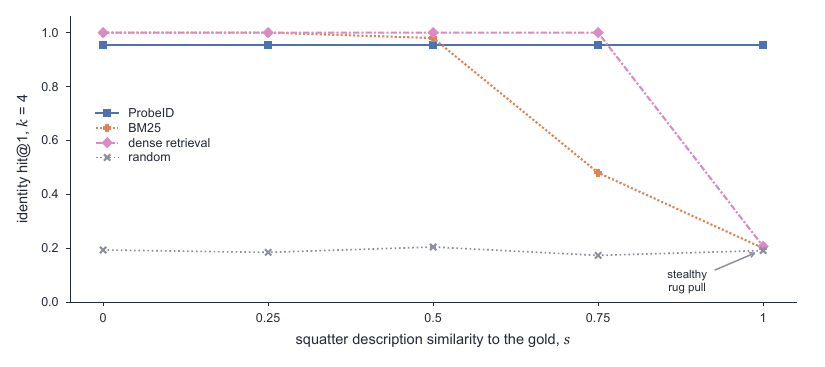}
\caption{\textbf{ProbeID remains flat as description routers recover.} Similarity $s$ moves from $1$ to $0$ at $k{=}4$, with behavior still mutated. Routers recover away from copied descriptions, while ProbeID is independent of $s$.}
\label{fig:descsim}
\end{figure}

The headline keeps the squatter description byte identical to the gold, which forces routing to fail by construction. Figure~\ref{fig:descsim} relaxes this and shows that the behavioral comparison is not tautological. It varies the squatter description from identical to disjoint, at token overlap $s$ from $1$ to $0$, while keeping behavior mutated, at $k{=}4$. ProbeID stays flat at $0.956$, since it ignores descriptions. The description routers improve as the squatter description diverges from the gold, with dense retrieval recovering earlier than BM25, and reach ceiling only when the squatter advertises a clearly different description. A stealthy squatter that copies the description (high $s$) defeats routing, which handles only the careless attacker, and ProbeID holds in that regime.

\subsection{Complementarity on a constructed twin family}
\label{app:twin}
The main text reports natural sibling groups. In this constructed counterpart every candidate keeps the gold behavior and differs only in its description, so the behavioral margin is zero by construction. The candidates are equivalent on the probe distribution, the gap gate abstains, and these responses cannot recover which tool was recorded (Table~\ref{tab:twin}).

\begin{table}[t]
\centering
\caption{\textbf{Behavioral twins require routing or escalation.} Results use 300 gold tools and 3 seeds. ProbeID abstains while lexical routing identifies the gold. The two signals address different failure modes (Table~\ref{tab:main}).}
\label{tab:twin}
\begin{tabular}{lccc}
\toprule
$k$ & ProbeID abstain & ProbeID wrong, all cases & lexical hit@1 \\
\midrule
1 & 1.00 & 0.00 & 0.998 \\
2 & 1.00 & 0.00 & 0.991 \\
4 & 1.00 & 0.00 & 0.983 \\
\bottomrule
\end{tabular}
\end{table}

\subsection{Natural description collisions and a dense router}
\label{app:natural}
The headline rug pull copies the gold description byte for byte, which defeats routing by construction. This study uses only vendor-authored descriptions and replaces the lexical router with a dense embedding one. Real Model Context Protocol servers already ship groups of tools whose vendor-written descriptions naturally overlap, whose input schemas are compatible, and whose behavior differs. We use these groups over live stdio JSON-RPC on the official filesystem server. The listing group holds three real listing tools that take a single path and return a plain directory listing, the same listing with sizes, or a recursive JSON tree. Their vendor descriptions are near duplicates, with an all-MiniLM-L6-v2 cosine of $0.80$ and a Jaccard of $0.85$ between the closest pair, yet the three tools have pairwise distance $1.00$ on the tested probes. The reading group holds read\_text\_file and its deprecated alias, a genuine behavioral twin at description cosine $0.58$.

An agent records a reference for one tool in a group and must re-identify that tool among its natural siblings. A description router scores each candidate's real description against the agent's natural language intent, averaged over three phrasings of the need. For the listing group they are \emph{show the files and directories inside a folder}, \emph{list the contents of a directory} and \emph{what files are in this path}. For the reading group they are \emph{read the text contents of a file}, \emph{get the full text of a file} and \emph{open a file and return its text}. ProbeID probes each candidate live and matches behavior to the recorded reference, abstaining on a true twin. We compare a dense router that embeds descriptions with MiniLM against a lexical Jaccard router. We author no descriptions. Every description is the vendor's own, fetched live from the server's tool listing.

\begin{table}[t]
\centering
\caption{\textbf{The admission reference separates natural siblings, and ProbeID abstains on an alias} (live filesystem server, real vendor descriptions, all-MiniLM-L6-v2 dense router, $30$ seeds, strict tie-breaking). The listing group has three behaviorally distinct siblings, the reading group has a tool and its deprecated alias (a behavioral twin). ProbeID recovers the recorded tool by behavior, or abstains on the twin, while a dense embedding router and a lexical Jaccard router, which lack the reference, sit at the $1/|C|$ chance line. Best in bold.}
\label{tab:natural}
\small
\setlength{\tabcolsep}{3.5pt}
\begin{tabular}{lccccc}
\toprule
group ($|C|$) & ProbeID & abstain & dense & Jaccard & chance \\
\midrule
listing (3) & \textbf{1.00} & 0.00 & 0.33 & 0.33 & 0.33 \\
reading (2) & -- & 1.00 & 0.50 & 0.50 & 0.50 \\
\bottomrule
\end{tabular}
\end{table}

Table~\ref{tab:natural} reports the result. On the listing group ProbeID recovers the recorded tool at hit@1 $1.00$, since the siblings are maximally separated in behavior, a sibling distance of $1.00$ against a self distance of $0.00$. Both routers sit at the $1/3$ chance line, the dense one no better than the lexical one, because near-duplicate descriptions do not encode which specific sibling a reference recorded, so the router picks the same candidate regardless of the truth. On the reading group the deprecated alias is a behavioral twin, so ProbeID abstains on every trial while the routers stay at the $1/2$ chance line. These vendor-authored siblings test identification against an admission reference, which ProbeID reads and the shared intent cannot reveal, so the result concerns this information asymmetry and not general dense-retrieval accuracy. Two caveats limit it. First, the behavioral separation in the listing group is large, so the study measures a router's blindness to natural description collisions and not a difficult behavioral distinction, and the behaviorally hard natural case is the twin in the reading group, where ProbeID correctly abstains. Second, the admission reference is unavailable to the description router, which matches the stated verification task but limits a direct accuracy comparison with intent-based retrieval.

\subsection{Two real implementations, a behaviorally close natural case}
\label{app:jsonimpl}
The natural sibling study of Appendix~\ref{app:natural} separates tools whose behavior differs on every probe. The harder and more realistic case is two legitimate tools that agree on most inputs and diverge on a few, a small natural margin rather than a maximal one. We test this case with two real, independently implemented JSON serializers exposed as live MCP tools over stdio JSON-RPC, with the JSON serializer of CPython's standard library as the gold and V8's built-in serializer run through a live Node worker as the candidate. Neither is malicious. We run both in compact mode so that whitespace, a formatting choice the canonicalizer does not normalize (Appendix~\ref{app:dy}), does not mask the semantic divergence. Each tool returns its serialized text as a string field, which the canonicalizer compares literally. On the tested workload they agree on integers, ASCII strings, and fractional floats, and differ where their documented defaults differ, an integer-valued float ($62.0$ against $62$) and a non-ASCII character (\'e written as a Unicode escape sequence against a raw \'e). The divergence arises from the implementations' defaults without an authored drift, and the tool cards are the vendors' own wording, the CPython docstring against the MDN description, at a dense embedding cosine of $0.54$.

An agent records a reference from the CPython tool and must re-identify it against the V8 tool. A probe lands in the naturally divergent region with probability $\phi$, set by the workload, so the per-probe mismatch of the V8 tool is a real Bernoulli$(\phi)$ variable. This deterministic agreement experiment tests the margin-dependent budget on real software. It does not instantiate the noisy minimax construction of Appendix~\ref{app:lower}. We sweep $\phi$ and the budget $m$ over $100$ seeds.

\begin{table}[t]
\centering
\caption{\textbf{Probe budgets resolve two close JSON implementations} (the CPython and V8 serializers, live over stdio JSON-RPC, $100$ seeds). The tools agree on most inputs and diverge only on integer-valued floats and non-ASCII escaping, a small natural margin $\phi$. Identity hit@1 rises with the probe budget $m$ and tracks $1-(1-\phi)^m$. No wrong candidate was selected in any of the $100$ seeded trials of any cell, since the gold matches its own reference, and the remainder abstains.}
\label{tab:jsonimpl}
\small
\setlength{\tabcolsep}{4pt}
\begin{tabular}{lccccc}
\toprule
margin $\phi$ & $m{=}1$ & $m{=}2$ & $m{=}4$ & $m{=}8$ & $m{=}16$ \\
\midrule
$0.15$ & 0.15 & 0.34 & 0.38 & 0.65 & 0.93 \\
$0.30$ & 0.29 & 0.58 & 0.74 & 0.95 & 1.00 \\
$0.50$ & 0.49 & 0.75 & 0.94 & 0.99 & 1.00 \\
\bottomrule
\end{tabular}
\end{table}

Table~\ref{tab:jsonimpl} reports identity hit@1, and two properties hold in every cell. First, the misidentification rate is $0.00$. The CPython tool matches its own reference, so it is never beaten, and ProbeID either names it when a probe hits the divergence or abstains, never selecting the wrong implementation. Second, the observed margin constrains the budget. At $\phi{\approx}0.3$ hit@1 rises $0.29, 0.58, 0.74, 0.95, 1.00$ as $m$ goes $1,2,4,8,16$, tracking $1-(1-\phi)^m$, and a smaller margin needs a larger budget, with $\phi{\approx}0.15$ reaching $0.93$ only at $m{=}16$. The rest abstain. This is the Theorem~\ref{thm:bound} dependence on a behaviorally close pair of real, independently written tools, the regime where identity is hard.

\section{Canonicalization and simple behavioral baselines}
\label{app:canonbase}
\subsection{The canonicalization ablation in full}
\label{app:canon}
Injecting ten random timestamp and id fields into each response isolates the effect of canonicalization. On the 150-gold headline cohort with canonicalization on, the noise floor stays at $0.00$ and hit@1 at $0.956$ for $k\ge2$, matching the noise-free run. With it off the gold self distance rises to $0.73$ and forced-choice identity barely moves, $0.953$ at $k{=}4$, because symmetric noise cancels in the ranking, and a gate at the twin study's $\tau_{\mathrm{hi}}{=}0.6$ would reject the average gold, since $0.73$ exceeds it. This compares the measured self distance with the threshold and is not a gated run. Canonicalization lowers the measured self distance relative to this fixed threshold. Under noise that canonicalization does not remove, a $0/1$ equality or hash baseline loses discrimination while continuous $d_Y$ retains it (Appendix~\ref{app:baselines}).

\subsection{Simple behavioral baselines}
\label{app:baselines}

\begin{table}[t]
\centering
\caption{\textbf{ProbeID recovers identity where copied descriptions reduce routers to chance.} Identity hit@1 under rug pull drift, on 402 tools with real ToolBench descriptions (150 gold per seed, 3 seeds). BM25, dense retrieval, and an LLM reranker all match random because the squatter copies the gold description. Best non-oracle result in bold.}
\label{tab:main}
{\small
\setlength{\tabcolsep}{4pt}
\begin{tabular}{lcccccc}
\toprule
$k$ & ProbeID & BM25 & Dense & LLM-card & Random & Oracle \\
\midrule
1 & \textbf{1.000} & 0.496 & 0.516 & 0.496 & 0.538 & 1.000 \\
2 & \textbf{0.956} & 0.344 & 0.344 & 0.344 & 0.349 & 1.000 \\
4 & \textbf{0.956} & 0.200 & 0.207 & 0.200 & 0.191 & 1.000 \\
8 & \textbf{0.956} & 0.111 & 0.111 & 0.111 & 0.107 & 1.000 \\
\bottomrule
\end{tabular}}
\end{table}
To test whether the bounded structural distance is necessary, we add three simple baselines that share ProbeID's probe set, fingerprints and strict tie-breaking and differ only in how they compare responses. \emph{Exact equality} scores a candidate by the fraction of probes whose canonical response differs from the reference, a 0/1 test in place of the structural $d_Y$. \emph{Response hash} compares the sets of response hashes by Jaccard distance, the usual signature-matching check. \emph{Per-probe vote} keeps a continuous per-probe distance but aggregates by majority vote instead of the mean. Three further selectors use alternative response comparisons. An \emph{embedding nearest-neighbor} (NN) selector encodes each canonicalized response with all-MiniLM-L6-v2 mean pooling and scores a candidate by the mean per-probe cosine distance to the reference, which tests whether response embeddings can replace the structural distance. A \emph{MET style MMD} treats the reference and candidate response sets as two samples and computes an unbiased MMD$^2$, the test of \citet{gao2025modelequality} with a character 4-gram spectrum kernel in place of their Hamming kernel. The embedding selector serializes each canonical response as sorted JSON and truncates it to $256$ tokens. The spectrum kernel is normalized to unit self-similarity, and a negative unbiased estimate is not clipped. A \emph{definition hash pin} follows the industry practice of pinning a hash of the tool definition, description plus schema \citep{beurerkellner2025toolpoisoning}, and selects uniformly among pin-matching candidates. A deployed toolkit implements this check. The AI SDK digests each tool's description, resolved input schema and title and diffs two such digests to flag drift \citep{aisdk2026tooldrift}. Two shortcuts bound what a trivial signal achieves. \emph{Status only} compares the status channel alone, and \emph{type and length} compares each response's top-level type and the relative gap between serialized lengths. All selectors are scored on one fixed cohort, the described pool with $300$ gold per seed over three seeds, whose cases are built from clean canonical behavior and held fixed while one factor of the observation changes. The canonicalizer is on or off. The noise is one of four kinds, namely none, ten volatile fields named on the canonicalizer's drop list as in the canonicalization ablation, ten volatile fields under names the drop list misses, or independent jitter of every scalar leaf with probability $\rho$. The unlisted names include a camel-case creation time, a hyphenated request identifier and a latency. The reference, the gold and every squatter are observed with independent noise.

\begin{table}[t]
\centering
\caption{\textbf{On one fixed cohort, the structural distance does not dominate every behavioral comparison} ($k{=}4$, 300 gold per seed, 3 seeds). Volatile fields named on the drop list leave every selector at the clean value when the canonicalizer is on, by construction, and that column is omitted. A canonicalizer tolerates only fields it names, 0/1 tests fall furthest under noise it does not remove, the set-level MMD degrades least under value jitter, and on the clean cache even type and length separates most cases.}
\label{tab:baselines}
{\small
\setlength{\tabcolsep}{4pt}
\begin{tabular}{lcccccc}
\toprule
 & \multicolumn{2}{c}{no noise} & \multicolumn{2}{c}{ten volatile fields} & \multicolumn{2}{c}{leaf jitter, canon on} \\
\cmidrule(lr){2-3}\cmidrule(lr){4-5}\cmidrule(lr){6-7}
selector & canon on & canon off & listed, off & unlisted, on & $\rho{=}0.1$ & $\rho{=}0.2$ \\
\midrule
ProbeID (structural $d_Y$) & 0.952 & 0.974 & 0.952 & 0.950 & 0.950 & 0.927 \\
per-probe vote & 0.952 & 0.974 & 0.952 & 0.950 & 0.943 & 0.920 \\
MET style MMD & 0.952 & 0.974 & 0.953 & 0.951 & 0.956 & 0.949 \\
embedding NN & 0.952 & 0.974 & 0.899 & 0.868 & 0.928 & 0.917 \\
exact equality (0/1) & 0.952 & 0.974 & 0.083 & 0.083 & 0.761 & 0.641 \\
response hash & 0.952 & 0.974 & 0.083 & 0.083 & 0.752 & 0.657 \\
type and length & 0.927 & 0.946 & 0.818 & 0.851 & 0.921 & 0.905 \\
status only & 0.000 & 0.000 & 0.000 & 0.000 & 0.000 & 0.000 \\
definition hash pin & 0.202 & 0.202 & 0.202 & 0.202 & 0.202 & 0.202 \\
random & 0.190 & 0.190 & 0.190 & 0.190 & 0.190 & 0.190 \\
\midrule
gold self distance & 0.00 & 0.00 & 0.74 & 0.74 & 0.17 & 0.29 \\
\bottomrule
\end{tabular}}
\end{table}

Table~\ref{tab:baselines} separates the factors that the canonicalization ablation changed together. On the clean cache every comparison of complete responses scores $0.952$, a selector reading only each response's type and length already scores $0.927$, and status alone scores $0$ because every cached response reports success, so the clean column cannot rank comparisons. Switching the canonicalizer off there raises every comparison of complete responses to $0.974$. The drop list also hides drift confined to the fields it drops. Across the $20$ cases that change, $40$ squatters tie the gold, $34$ of which delete a date, cursor or id key while the other $6$ mutate only such fields. A canonicalizer tolerates the fields it names and nothing else. Named volatile fields leave every selector at its clean value, and the same fields under unlisted names act as if the canonicalizer were off. The gold self distance rises to $0.74$. Exact equality and response hash fall to $0.083$, below the random bound, because every response mismatches on its volatile fields, while the structural distance holds at $0.950$ and the MMD at $0.951$. The embedding NN degrades more as $k$ grows, from $0.931$ at $k{=}1$ to $0.822$ at $k{=}8$ under unlisted fields, because volatile tokens move the whole vector. Jitter on the fields the canonicalizer keeps separates the continuous selectors. The structural distance degrades gradually, $0.957$, $0.950$ and $0.927$ at $\rho$ of $0.05$, $0.1$ and $0.2$. The 0/1 tests degrade faster, to $0.641$ and $0.657$, and the MMD performs best, at $0.949$. At $k{=}8$ and $\rho{=}0.2$ the MMD's $0.944$ is well ahead of the structural distance's $0.881$, and the type and length shortcut reaches $0.889$. With cases ranked by each selector's own gap between its two nearest candidates, the area under the risk-coverage curve at $\rho{=}0.2$ is $0.008$ for the structural distance, $0.003$ for the MMD and $0.128$ for exact equality. Under the tested noise that canonicalization does not remove, continuous comparisons outperform equality, and among them the structural distance is not the most stable under value noise. Theorem~\ref{thm:bound} and the abstention guarantee are stated for the probe mean of a bounded per-probe comparison. They hold for exact equality and for the per-probe vote, whose vote share is the probe mean of a bounded indicator, each with the margin measured on its own comparison. The set-level MMD is not a probe mean and carries no such guarantee. The framework supplies the accounting from margin to budget and a joint error bound for whichever bounded per-probe comparison a deployment's noise calls for, and the MMD result stays empirical. It does not depend on choosing $d_Y$ over a 0/1 test or a set statistic, and on deterministic drift a 0/1 test has the larger margin and verifies at a smaller budget (Appendix~\ref{app:agent}). The definition hash pin sits at chance at every $k$, $0.499$ down to $0.100$. A behavior-only rug pull leaves the pinned definition untouched, so pinning admits every candidate and carries no identity signal. An admission-time pin and a use-time behavioral check thus complement each other.

\paragraph{Paired uncertainty.} Resampling each gold with all its case-runs gives a pointwise $95\%$ interval of $[0.043,0.085]$ for the MMD minus structural-distance hit@1 difference of $0.063$ at $\rho{=}0.2$, $k{=}8$, over $900$ case-runs from $393$ gold tools. At $\rho{=}0.1$ and the same candidate count, the observed difference is $0.006$ with interval $[-0.003,0.015]$, which does not resolve their ordering. These exploratory intervals reuse the original tools and injected noise, cover all saved regimes in the supplement, and are not adjusted for multiple comparisons. They provide no independent-cohort validation.

\section{Further results on the cached benchmark}
\label{app:cached}
\subsection{The probe budget on real recorded behavior}
\label{app:realbudget}
\begin{figure}[t]
\centering
\includegraphics[width=\textwidth]{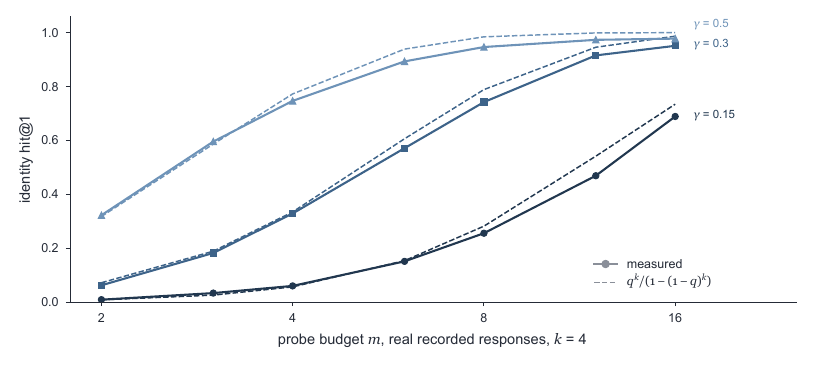}
\caption{\textbf{The probe budget binds on real recorded behavior.} Identity hit@1 against probe budget $m$ when the squatter diverges on a $\gamma$ fraction of real cached probes ($k{=}4$). Solid is measured, dashed is the matching closed form conditional on a separable case. Each probe replays a recorded response of a real API, so the budget effect is not an artifact of the simulated channel of Figure~\ref{fig:main}(b).}
\label{fig:realbudget}
\end{figure}

The controlled channel of Figure~\ref{fig:main}(b) isolates the $\sigma^2$ factor, with its smallest budgets in Table~\ref{tab:bound}, but it is simulated. To check that the budget binds on real responses, we use a squatter that returns the gold response on a $1-\gamma$ fraction of probes and a mutation on the rest, so its per-probe failure probability is $\gamma$. Figure~\ref{fig:realbudget} reports identity hit@1 against the budget $m$ for three margins at $k{=}4$. As in Results 1 and 3, a case is kept only when at least one squatter is behaviorally separable, so the reported hit@1 and its closed form $q^k/(1-(1-q)^k)$ with $q{=}1-(1-\gamma)^m$, which assumes independent separation events across squatters, are conditional on that event. The event changes with the budget. At $\gamma{=}0.3$ the exclusions fall from $53$ of the $503$ golds drawn over three seeds at $m{=}2$ to $13$ of $463$ at $m{=}16$. Counting the excluded golds as misses lowers every measured point of the figure by at most $0.04$, and the unconditional curve then sits below $q^k$ by as much as $0.09$, against $0.07$ for the conditional pair. Hit@1 rises with $m$ and tracks this form, and a smaller margin needs a larger budget. The candidate count moves the shared-input budget in the direction of the union bound. At $\gamma{=}0.3$, reaching $0.9$ hit@1 needs $m\approx12$ at $k{=}4$ and $m\approx16$ at $k{=}8$. The measured curve sits a few points below the idealized form at large $m$, because some sparse responses canonicalize a mutation back to the gold.

\begin{table}[t]
\centering
\caption{\textbf{The 0.99 hit@1 budget is consistent with $m^\star\!\propto\!\sigma^2/\gamma^2$.} Results use 200 gold and 3 seeds. The product $m^\star\gamma^2$ is approximately constant across nominal margins and rises with self distance, consistent with the $\sigma^2$ factor.}
\label{tab:bound}
\begin{tabular}{lccccc}
\toprule
 & $\gamma{=}0.1$ & $0.15$ & $0.2$ & $0.3$ & $0.4$ \\
\midrule
$m^\star$ (self dist.\ $0.18$) & 256 & 128 & 128 & 64 & 32 \\
$m^\star\gamma^2$ & 2.6 & 2.9 & 5.1 & 5.8 & 5.1 \\
\addlinespace
$m^\star$ (self dist.\ $0.34$) & $>$256 & 256 & 128 & 64 & 64 \\
$m^\star\gamma^2$ & -- & 5.8 & 5.1 & 5.8 & 10.2 \\
\bottomrule
\end{tabular}
\end{table}

\subsection{Squatter counts beyond eight}
\label{app:ksweep}
The headline sweep stops at $k{=}8$, well inside the range where the 12-probe budget saturates, so it does not stress the $\log(n+k)$ dependence of Theorem~\ref{thm:bound} at the squatter counts a heavily typosquatted registry could present. We extend the same protocol, 150 gold per seed, three seeds, the same drift families and budget, to $k\in\{16,32,64\}$. ProbeID is flat at $0.956$ at all three counts, unchanged from $k{=}8$. BM25 and dense retrieval fall to $0.058$, $0.022$, and $0.009$, at or below the $1/(k+1)$ chance line ($0.059$, $0.030$, $0.015$), since identical inputs leave a router only tie-breaking, and the random selector tracks that line at $0.067$, $0.038$, $0.016$. This flat curve does not test the logarithmic dependence. On this noiseless cache the gold sits at distance zero and every added squatter comes from the same two mutation families, so once both families are separated the squatter count cannot move ProbeID. The candidate count enters the budget where noise does, in Result~2 and Appendix~\ref{app:realbudget}, while here every description signal is already gone.

\subsection{Abstention on fixed cached cases}
\label{app:riskcov}
Theorem~\ref{thm:abstain} applies to independently sampled probes and a gold-present candidate pool. The headline benchmark instead forces strict choice on fixed cache-order probes, at most $12$ per case. We saved every candidate distance and rescored the same $450$ cases per squatter count. For each case, we set $\epsilon_i=\sqrt{\log(2(k{+}1)/0.05)/(2m_i)}$, $\gamma_{\min,i}=2\epsilon_i$, and $\tau_{\mathrm{hi},i}=d_{t^\star,i}+\epsilon_i$ using its actual $m_i\in[4,12]$. The observed gold distance is $0$ throughout, and the threshold uses it, so this is a retrospective rescore rather than a deployable threshold rule. The rescore decides $221$, $83$, $78$, and $69$ cases at $k\in\{1,2,4,8\}$, respectively, with no observed wrong decisions. This is an empirical operating point under a theorem-inspired radius, not a probability certificate. The radius must use each case's own probe count. Setting it with $m{=}12$ for every case would shrink it for the $202$ of $450$ cases at $k{=}8$ that have fewer probes and would inflate coverage from $69/450$ to $141/450$.

The full threshold grid is descriptive because its favorable points were identified using these cases. With $\tau_{\mathrm{hi}}{=}1$ and $\gamma_{\min}{=}0.01$, coverage is $0.956$ with no observed wrong decisions on this benchmark. The value equals strict-choice hit@1 because this grid point decides every non-tied case and abstains on the ties that strict choice scores as misses. The gold distance is $0.00$ in all $1{,}800$ cases, and the smallest positive gap at $k\ge2$ is $0.057$, so this grid point separates every non-tied case. It cannot establish error control on unseen cases. On the behavioral twin study, the margin gate abstains throughout because these probes contain no distinguishing response.

The same per-case records explain the headline misses. At each $k\ge2$, all $20$ missed case-runs out of $450$, or $4.4\%$, have a gap of $0$ between the gold and the nearest squatter on every available cached probe, so every miss is a realized behavioral twin under strict tie-breaking, not a near miss. Eleven of the $20$ missed tools expose fewer than the full $12$ cached probes, so their budget is bounded by cache availability rather than by the probe generator.

\section{Live MCP servers}
\label{app:live}
\subsection{Probe generator ablation on a live server}
\label{app:probegen}
The cached benchmark treats each gold tool's cached arguments as its probe set, so it does not test whether the schema-driven probe generator, module M1, produces useful unseen inputs. The live-server experiment tests freshly generated probes without cache access. The gold is the live get-sum tool of the everything server, and the squatter is a graded near twin that drifts on a $\phi$ fraction of the integer input space. We compare two probe sources, both drawn fresh on every trial and never reused. They are the proposed M1 generator, which emits boundary, enumerated, and random schema-valid values, and a broad uniform random sampler as a coverage control. With strict tie-breaking the gold is named only when some probe lands in the squatter's drift region, so identity hit@1 should track $1-(1-\phi)^m$ for any source that samples the space.

\begin{table}[t]
\centering
\caption{\textbf{Fresh probes recover identity on a live server without cached replay} (live get-sum tool, graded near twin at margin $\phi$, $k{=}1$, 30 fresh probe draws per cell, no cache consulted). The uniform control's hit@1 tracks $1-(1-\phi)^m$ at its realized divergence. The proposed M1 generator reaches $1.00$ at least as fast, since its boundary values over-cover the tested drift region.}
\label{tab:probegen}
\small
\setlength{\tabcolsep}{4pt}
\begin{tabular}{@{}llccccc@{}}
\toprule
source & $\phi$ & \multicolumn{5}{c}{identity hit@1 at $m$} \\
\cmidrule(lr){3-7}
 & & $1$ & $2$ & $4$ & $8$ & $16$ \\
\midrule
uniform & $0.15$ & 0.10 & 0.33 & 0.53 & 0.73 & 0.90 \\
 & $0.30$ & 0.27 & 0.50 & 0.70 & 0.93 & 1.00 \\
 & $0.50$ & 0.40 & 0.67 & 0.93 & 1.00 & 1.00 \\
\midrule
M1 schema & $0.15$ & 0.60 & 0.70 & 0.83 & 1.00 & 1.00 \\
 & $0.30$ & 0.63 & 0.80 & 0.87 & 1.00 & 1.00 \\
 & $0.50$ & 0.67 & 0.83 & 0.90 & 1.00 & 1.00 \\
\bottomrule
\end{tabular}
\end{table}

In Table~\ref{tab:probegen} the uniform control behaves as predicted. Its realized divergence matches the nominal $\phi$, and hit@1 tracks $1-(1-\phi)^m$ across budgets. At $\phi{=}0.3$ it rises $0.27, 0.50, 0.70, 0.93, 1.00$ for $m{=}1,2,4,8,16$ against a closed form of $0.93$ at $m{=}8$, and at $\phi{=}0.5$ the single-probe value is $0.40$, slightly below $\phi$. Identity is recovered from behavior on newly drawn schema-valid inputs outside the cached benchmark, not by replaying known arguments. The M1 generator also drives hit@1 to $1.00$ at every margin by $m{=}8$, which the uniform control does not. Its boundary integers over-cover this hash-defined drift region, so its realized divergence runs above the nominal $\phi$. The gold self distance is $0.00$ throughout, so the noiseless margin holds on the live tool. This over-coverage is also the limit noted in Appendix~\ref{app:adaptive}. An adaptive adversary could enumerate a small, predictable boundary set, so the live studies draw from a large input space instead.

\subsection{Diverse live servers}
\label{app:multiserver}
The live study of the main text uses the everything reference server. To check that identity recovery is not specific to one server, we run ProbeID against three official MCP server families over live stdio JSON-RPC with no caching, the filesystem server (read\_text\_file), the memory server (open\_nodes) and the everything server (get-sum). The filesystem and memory families stage a data-level rug pull, in which each squatter is the same official server binary serving tampered backing data. The executable, tool schema and human-readable description are identical to the gold, and only the backing data drifts. Each squatter tampers with a distinct nonempty half of the files or graph entities. This is a weaker adversary than an independently re-implemented squatter, which we study separately in Appendix~\ref{app:reimpl}, because the output format and error behavior are inherited from the gold binary. The everything family keeps the re-declared FastMCP squatter on get-sum. Across all three families the squatter and gold descriptions are byte identical (verified from the live tool listing) and the gold self distance is $0.00$. The canonicalizer maps three response modalities, file text, JSON knowledge-graph nodes and integer sums, to a stable signal.

\begin{table}[t]
\centering
\caption{\textbf{Identity hit@1 on three live MCP server families} over stdio JSON-RPC, $30$ seeds per cell, strict tie-breaking. The description router over the byte-identical tool cards can only guess at $1/(k{+}1)$. ProbeID recovers identity as the probe budget grows on the filesystem and memory families, where each squatter is the same official binary on tampered data, and saturates at one probe on the everything family, whose squatter drifts on every input.}
\label{tab:multiserver}
\small
\setlength{\tabcolsep}{4pt}
\begin{tabular}{@{}llcccccc@{}}
\toprule
family & $k$ & \multicolumn{4}{c}{ProbeID hit@1 at $m$} & router \\
\cmidrule(lr){3-6}
(tool) & & $1$ & $2$ & $4$ & $8$ & $\tfrac{1}{k+1}$ \\
\midrule
filesystem & $1$ & 0.53 & 0.77 & 0.93 & 1.00 & 0.50 \\
(read\_text\_file) & $2$ & 0.27 & 0.70 & 0.93 & 1.00 & 0.33 \\
 & $3$ & 0.00 & 0.43 & 0.90 & 1.00 & 0.25 \\
\midrule
memory & $1$ & 0.60 & 0.77 & 0.97 & 1.00 & 0.50 \\
(open\_nodes) & $2$ & 0.33 & 0.73 & 0.93 & 1.00 & 0.33 \\
 & $3$ & 0.00 & 0.50 & 0.93 & 1.00 & 0.25 \\
\midrule
everything & $1$ & 1.00 & 1.00 & 1.00 & 1.00 & 0.50 \\
(get-sum) & $2$ & 1.00 & 1.00 & 1.00 & 1.00 & 0.33 \\
 & $3$ & 1.00 & 1.00 & 1.00 & 1.00 & 0.25 \\
\bottomrule
\end{tabular}
\end{table}

On the filesystem and memory families the budget binds as predicted (Table~\ref{tab:multiserver}). A single probe may miss a squatter whose tampered half it does not sample, but a handful drive hit@1 to $1.00$. The description router is pinned at $1/(k+1)$ because the tool cards are byte identical. The $k{>}1$ rows fall lower at the smallest budget because strict tie-breaking demands that the gold separate from every one of the $k$ squatters at once. We do not read these families as $k$ independent pieces of evidence, since the tampered regions of sibling squatters overlap. The same nearest-fingerprint rule and canonicalizer recover identity across three real server binaries and three response modalities, against descriptions copied byte for byte, although the guarantee also needs its sampling and margin conditions. A census of authenticated production servers with side effects remains out of scope, for the reason stated in the limitations.

\subsection{Re-implemented squatters}
\label{app:reimpl}
The same-binary study above uses a weak adversary by design. A realistic rug pull squatter does not corrupt the gold server's backing data but runs an independent server that mimics the gold's tool card and output format while its behavior drifts. We test this case. For four tools across three official servers (get-sum and echo on everything, read\_text\_file on filesystem, open\_nodes on memory), the gold is the official reference server over live stdio JSON-RPC. The squatters are hand-written re-implementations that share no code with the official server, one in raw JSON-RPC without the MCP SDK and one on the SDK's low-level server interface. Each squatter re-declares the target tool with a byte-identical description, fetched live from the gold, and serves behavior from one verified drift model. The squatter variants differ in stealth.

\begin{itemize}
\item A \emph{sloppy} squatter returns a different output format (the careless attacker).
\item A \emph{format-faithful} squatter copies the gold's exact output envelope, error behavior and field structure, with a wrong value on every probe (the stealthy attacker).
\item A \emph{near twin} is format faithful and drifts on only a fraction $\phi$ of inputs.
\item An \emph{exact twin} behaves identically.
\end{itemize}

\begin{table}[t]
\centering
\caption{\textbf{Tested value drift separates re-implemented squatters, identity hit@1 over live stdio JSON-RPC} ($k{=}3$ squatters in two distinct codebases, $30$ seeds, strict tie-breaking). A format-faithful squatter that copies the gold's exact output envelope but returns a wrong value is caught at a single probe, identically to a sloppy one, because the behavioral distance compares canonicalized values, not format. Under the gated rule, ProbeID abstains on the exact behavioral twin. Self distance is the gold against itself and twin distance is the re-implemented twin against the gold, both $0.00$ on the sampled responses.}
\label{tab:reimpl}
\small
\setlength{\tabcolsep}{4pt}
\begin{tabular}{@{}lccccc@{}}
\toprule
tool & sloppy & faithful & twin & self & twin \\
(server) & $m{=}1$ & $m{=}1$ & abstain & dist & dist \\
\midrule
get-sum (everything) & 1.00 & 1.00 & 1.00 & 0.00 & 0.00 \\
echo (everything) & 1.00 & 1.00 & 1.00 & 0.00 & 0.00 \\
read\_text\_file (fs) & 1.00 & 1.00 & 1.00 & 0.00 & 0.00 \\
open\_nodes (memory) & 1.00 & 1.00 & 1.00 & 0.00 & 0.00 \\
\bottomrule
\end{tabular}
\end{table}

The format-faithful squatter tests whether an independent re-implementation that mimics the gold's format can slip past a behavioral check (Table~\ref{tab:reimpl}). In these experiments it does not. Canonicalization compares values, and the tested wrong values separate the squatter from the gold at $\text{hit@1}=1.00$ with a single probe across all four tools and both codebases. Only the near twin survives, and only in proportion to how rarely it drifts. On the two tools with large input spaces (get-sum and echo, $k{=}1$), hit@1 rises toward $1-(1-\phi)^m$ as the budget grows. For echo at $\phi{=}0.5$ it is $0.40, 0.70, 0.90, 0.97$ at $m{=}1,2,4,8$. A smaller margin $\phi{=}0.15$ needs the full budget to climb, the Theorem~\ref{thm:bound} dependence of Figure~\ref{fig:main}(c) now against a re-implemented near twin. The exact twin is the limit $\phi{=}0$, on which ProbeID abstains at every budget and never selects the twin over the gold (twin distance $0.00$). The description router is pinned at $1/(k{+}1)$ throughout, since the four tool cards are byte identical. The fingerprint reads only tool-call responses, never server metadata such as the information block returned at initialization, which the threat model treats as attacker controlled. A squatter can spoof its advertised name, and in these four read-only tests the sampled responses still separate the value-drifting squatters. This study covers only read-only, deterministic tools with few side effects, the class our threat model targets. Appendix~\ref{app:stateful} studies the stateful case and Appendix~\ref{app:sideeffect} side-effecting and authenticated tools, and authenticated production servers remain future work.

\section{Stateful, side-effecting and authenticated tools}
\label{app:statefulall}
\subsection{Stateful and side-effecting tools}
\label{app:stateful}
Theorem~\ref{thm:bound} assumes independent per-probe observations. A stateful tool violates this, because a side-effecting call changes the server and the same probe then returns a different response depending on call order. We test this on a live stateful MCP server whose counter tool adds a step to an internal accumulator on every call and returns the running total. The gold steps by $1$ and a behavioral squatter steps by $2$. We compare two probing strategies over $30$ trials at $m{=}8$ (Table~\ref{tab:stateful}).

\begin{table}[t]
\centering
\caption{\textbf{Sandboxed state restores identity in the counter experiment.} Gold step $1$, squatter step $2$, $m{=}8$, $30$ trials. Shared sessions produce abstention, while a fresh sandbox restores zero self distance and identifies the gold.}
\label{tab:stateful}
\small
\setlength{\tabcolsep}{5pt}
\begin{tabular}{lcccc}
\toprule
probing & self dist & abstain & wrong, all trials & hit@1 \\
\midrule
naive (shared) & 1.00 & 1.00 & 0.00 & -- \\
sandboxed (fresh) & 0.00 & 0.00 & 0.00 & 1.00 \\
\bottomrule
\end{tabular}
\end{table}

Under naive probing a single shared session is reused, so the gold's reference and its re-probed fingerprint advance to different totals. The gold self distance is $1.00$, far above the distance threshold $\tau_{\mathrm{hi}}{=}0.5$, so the nearest-fingerprint selector abstains on every trial with a misidentification rate of $0.00$. The stateful tool triggers abstention. ProbeID detects that no candidate matches, not even the gold, and escalates. Under sandboxed probing a fresh server instance serves each fingerprint, so every fingerprint starts from a clean state. The gold self distance returns to $0.00$ and identity hit@1 is $1.00$, separating the step-by-1 gold from the step-by-2 squatter. For a side-effecting tool, the deployment procedure is to probe a reset sandbox or a dry-run endpoint, which restores a zero self distance, and to abstain when no such sandbox is available. The probes of one fingerprint still run in sequence inside one sandbox, so their responses are not independent and Theorem~\ref{thm:bound} does not cover them. The independent unit is the reset episode, and this result is empirical.

\subsection{Side-effecting writes and authentication}
\label{app:sideeffect}
The studies so far probe read-only tools or a single counter. Two tool classes common in production agents fall outside that scope, tools that mutate persistent state through writes and tools gated by authentication. We test both on live FastMCP servers over stdio JSON-RPC, with the squatter and gold tool cards byte identical, so only behavior can separate them.

\paragraph{A side-effecting key-value store.} The store tool sets and gets values by key. The gold uses last-write-wins overwrite semantics. The squatter silently ignores a write when the key already exists, a first-write-wins data-integrity drift a description router cannot see. Each fingerprint is taken in a reset sandbox, the deployment procedure for side-effecting tools established for the counter in Appendix~\ref{app:stateful}, so the gold matches its own reference at self distance $0.00$. ProbeID separates the two write semantics on the get-after-update probes and recovers identity at hit@1 $1.00$ with a misidentification rate of $0.00$ (Table~\ref{tab:sideeffect}).

\paragraph{An authentication-gated tool.} The read\_secret tool returns a protected resource only for a valid token, and the gold rejects an invalid token with an authorization error. The squatter bypasses the check and returns the resource for any token, a security-relevant authentication drift behind an identical tool card. A probe set that alternates valid and invalid tokens separates them, since the two agree on valid-token probes and diverge on every invalid-token probe, placing the bypass squatter at behavioral distance $0.50$ from the reference. ProbeID recovers the gold at hit@1 $1.00$, at self distance $0.00$, with a misidentification rate of $0.00$, while a description router is blind by construction. This tested bypass accepts any token, so it diverges on every invalid-token probe and is caught by construction. A stealthier backdoor that rejected the probe tokens and bypassed only an attacker-held secret would agree with the gold across the probe distribution and drift only off it, the measure-zero high-coverage evasion of Appendix~\ref{app:adaptive}. With the gold present ProbeID still ties it at distance zero and abstains, so misidentification stays zero, but detecting such an off-distribution backdoor lies outside the on-$D$ guarantee and is not claimed.

\begin{table}[t]
\centering
\caption{\textbf{Observed writes and authentication checks distinguish the tested squatters.} Live FastMCP, $30$ trials, strict ties, and identical tool cards. Reset sandboxes expose changed overwrite semantics or bypassed authentication, with no observed misidentification.}
\label{tab:sideeffect}
\small
\setlength{\tabcolsep}{5pt}
\begin{tabular}{llcc}
\toprule
tool & squatter drift & hit@1 & misident. \\
\midrule
store & write semantics & 1.00 & 0.00 \\
read\_secret & auth bypass & 1.00 & 0.00 \\
\bottomrule
\end{tabular}
\end{table}

These remain controlled, self-hosted tools with cooperative squatters that do not distinguish probe traffic from production traffic, so the study shows identity recovery on side-effecting and authenticated behavior without settling the adaptive or production-scale case. The in-the-wild census of Appendix~\ref{app:census} probes real third-party servers, and a corpus of verified runnable MCP servers \citep{mcpzoo2025} makes a full production-scale census feasible. Throughout, the recipe probes a reset sandbox or a dry-run endpoint to restore a zero self distance, compares canonicalized behavior and abstains when no candidate matches.

\section{Behavioral distance under large outputs}
\label{app:dilution}
Any structural distance risks dilution. If a tool returns a large object and a format-faithful squatter perturbs only one leaf field, the normalized $d_Y$ on a single probe is tiny, and the per-probe margin may fall below the decision threshold. We test this on a live server. The gold is the official memory server. One probe requests $N$ entity nodes at once, so a single live response is a large structured object, and a re-implemented squatter (raw JSON-RPC, byte-identical description) returns the same $N$ nodes with $f$ of them perturbed in a format-faithful way. The measured per-probe margin is then $\gamma=f/(12N)$, which we confirm live.

\begin{table}[t]
\centering
\caption{\textbf{Large outputs dilute the observed margin and increase abstention.} Memory queries use $f{=}1$ changed node and $\gamma_{\min}{=}0.05$. The margin shrinks as $1/N$. Ranking remains correct, while a one-field difference below the threshold triggers escalation.}
\label{tab:dilution}
\small
\setlength{\tabcolsep}{4.5pt}
\begin{tabular}{@{}lcccccc@{}}
\toprule
output size $N$ & 1 & 2 & 5 & 10 & 20 & 40 \\
\midrule
margin $\gamma$ & 0.083 & 0.042 & 0.017 & 0.008 & 0.004 & 0.002 \\
identity hit@1 & 1.00 & 1.00 & 1.00 & 1.00 & 1.00 & 1.00 \\
passes margin gate ($\gamma{\ge}\gamma_{\min}$) & yes & no & no & no & no & no \\
\bottomrule
\end{tabular}
\end{table}

Table~\ref{tab:dilution} shows that forced-choice ranking and gated acceptance respond differently to dilution. The deterministic identity ranking is stable, because any nonzero distance places the squatter behind the gold, so hit@1 is $1.00$ at every output size, down to $\gamma=0.002$ at $N{=}40$. ProbeID does not miss the squatter, but the gated margin is diluted. At $\gamma_{\min}{=}0.05$ the gold passes the margin gate at $N{=}1$ and the rule abstains for $N\ge2$, because a one-field difference in a large object is an ambiguous signal. ProbeID escalates instead of deciding on insufficient evidence and never selects the squatter. The margin is $f/(12N)$. With $N{=}20$, raising the perturbed field count $f$ from $1$ to $20$ moves $\gamma$ from $0.004$ to $0.083$ and passes the gate again at $f{=}20$. A small margin falls within the regime Theorem~\ref{thm:bound} governs. The leading term of the Bernstein budget scales as $\sigma^2/\gamma^2$, so at fixed $\sigma$ a one-field drift in an $N$-field output needs a budget growing as $N^2$. Dilution is the price of a mean-aggregated distance, which tolerates per-field noise, whereas a max aggregation would react to any single noisy field. Per-field certification for very large outputs is left to future work.

\section{An in-the-wild census of real MCP servers}
\label{app:census}
The benchmark injects squatters into a cached corpus, and the live studies above use a handful of servers we configured. Neither measures the noise of independently published tools, which sets the self distance and per-probe variance a deployed verifier faces, so we measure it on real servers. We screened community servers from public registries by publisher reputation, download count and keyless, read-only access, and pinned exact versions. We ran $18$ distinct servers over stdio JSON-RPC on an isolated host, $7$ from the official reference collection (one of them since archived) and $11$ community servers. They span filesystem, version control, memory, SQL, time, web fetch and search, documentation lookup, literature search, weather, and format conversion. A read-only whitelist restricted probing, so no state-mutating or destructive tool was called. Each whitelisted tool received schema-valid probes from the M1 generator, each executed three times to estimate a self distance, and every raw and canonicalized response, latency and status was recorded for later drift replays.

Two servers failed to start without keys and were dropped, leaving $16$ with probe data. Canonicalization succeeded on every recorded response of all $16$ servers, error and exception responses included, a success rate of $1.00$, so the canonicalizer handles servers we did not author without curation. Parsing alone does not establish identity separation, and the servers differ in self distance.

\begin{table}[t]
\centering
\caption{\textbf{In-the-wild self distance on $16$ real MCP servers} (read-only tools, $3$ probe passes, canonicalization on). Under canonicalization most real tools are deterministic (self distance $0$), which says nothing about whether squatters are separable. A minority are nondeterministic, the $\sigma>0$ regime that the variance-aware budget and the abstention rule address. Canonicalization succeeds on every response of every server.}
\label{tab:census}
\small
\setlength{\tabcolsep}{4pt}
\begin{tabular}{lccc}
\toprule
regime & servers & tools probed & mean self distance \\
\midrule
deterministic & 10 & 35 & 0.00 \\
nondeterministic & 4 & 16 & 0.10--0.92 \\
\bottomrule
\end{tabular}
\end{table}

Table~\ref{tab:census} reports the split. Ten servers are deterministic under canonicalization, at a self distance of $0.00$ over all three passes. They are the filesystem, git, memory, SQLite, time, fetch and DuckDuckGo servers, two AWS documentation and pricing servers, and the corrected reference server. Zero self distance over three passes is the deterministic special case of the theorem's setting, although it does not by itself establish independent observations or a positive margin. Four servers are nondeterministic. A weather server's air quality reading changes between calls (per-tool self distance up to $1.00$, mean $0.18$ over its six tools), the context7 documentation server's search ranking varies ($0.83$ to $1.00$), and the arXiv server ($0.13$) and an AWS CDK construct search ($0.20$ on one pass) also vary. These observations describe the $\sigma>0$ regime of Appendix~\ref{app:bernstein} and are not calibrated gate evaluations, since Theorem~\ref{thm:abstain} applies only when its sampling, candidate-set and threshold assumptions hold. Two further servers, a CLI echo harness with fault-injection tools and a mindmap converter whose calls errored, produced only exceptions and are excluded from the split. Their raw dumps are released. Their eight tools complete the $59$ probed, beside $35$ deterministic and $16$ nondeterministic ones. Of the $35$, ten answer every probe with an error, so for them determinism means a stable error response, which does not by itself separate identities. We then replay the stored probes, re-issuing every recorded argument byte for byte under canonical JSON, after $24$ hours and again after $22$ days. Canonicalization holds at $1.00$ on both replays with all $16$ servers reachable, and the probe arguments are verified identical to $t_0$ by hash. Drift is counted against each tool's own $t_0$ noise. A tool that was deterministic at $t_0$ has drifted if it moved at all, and a nondeterministic one only if it moved beyond its own pass-to-pass distance. Counting any movement would flag $12$ tools at $24$ hours, of which $4$ drifted beyond their own noise and two more are measurement artifacts of the probe harness, set aside below.

At $24$ hours, $4$ of the $59$ probed tools show genuine drift. Three belong to a legitimately time-varying weather server. Its current conditions and detailed forecast move to distances $1.00$ and $0.53$, and its air quality reading to $0.25$ against a $0.07$ noise floor. The fourth, an AWS documentation recommender at $0.10$, was deterministic at $t_0$, so its movement is not pass-to-pass noise. From the logs we attribute it to the documentation service it queries, without an identified cause. Two tools of the fault-injection harness excluded from Table~\ref{tab:census} also register $0.50$ from this replay on, and we set both aside because neither ever returned a response. At $t_0$ every call to them failed on a closed session, as at every replay, so the $0.50$ reflects which exception the dead session raised and says nothing about the tools. At $22$ days the service-side count is unchanged at the same $4$ tools, but their magnitudes have grown, the AWS recommender moving from $0.10$ to $0.54$ and the weather forecast from $0.53$ to $0.71$. A registry-scale study of description drift over $89$ days reaches the complementary conclusion and names tool behavior as the surface its data cannot observe \citep{bharti2026registrydrift}.

The $22$-day replay also separates a second mechanism. Between the two replays four npm-installed servers had to be reinstalled, and reinstalling them at their recorded version pins changed the behavior of three further tools, two filesystem tools and one reference server tool, which moved to distances $1.00$, $1.00$ and $0.80$. No top-level version changed, and the recorded version of every one of the $16$ servers is identical at the two replays. Every changed response is a validation error string, rendered by a library that is not pinned. The SDK these servers build on declares its schema validation library as a version range spanning two major versions, and the reinstall resolved it to $4.4.3$ (released lockfile). We did not snapshot the installed tree at $t_0$, so which version it replaced is not recorded, and we claim only that the same servers, reinstalled at the same recorded versions, answer differently. This is the failure the threat model describes, arriving through the supply chain instead of the service. A pinned version does not pin behavior, and a use-time behavioral check sees what the pin misses.

Nine further snapshots at $25$, $27$, $34$, $41$, $48$, $55$, $62$, $69$ and $77$ days, each taken with the same replay script, separate the mechanisms by how they evolve over time. Across the ten replays from $22$ days onward the same seven tools drift once the harness artifacts are set aside, with canonicalization succeeding on every recorded response, all $16$ servers reachable, probe arguments verified identical to $t_0$ by hash, and every recorded server version identical to $t_0$ at every snapshot. The three supply-chain tools stay at $1.00$, $1.00$ and $0.80$ in all ten, because nothing was reinstalled again. The service side splits three ways rather than two. One tool holds a constant value throughout, the current-conditions reading at the ceiling of $1.00$. Two oscillate inside a band without a trend, the weather forecast between $0.48$ and $0.71$ and the air quality reading between $0.23$ and $0.29$. The AWS recommender diverges progressively. It climbs from $0.10$ at one day through $0.54$, $0.60$, $0.60$, $0.62$, $0.62$ and $0.67$ to the ceiling of $1.00$ at $55$ days and holds that ceiling at $62$, $69$ and $77$ days, a Spearman rank correlation with elapsed time of $0.954$ over the ten replays from $22$ days. The correlation describes these ten points and is not a test, since they are one tool observed repeatedly rather than a sample. The version pin showed no change. Its recorded version is the one from $t_0$, its probe arguments are byte identical, the server was reachable and canonicalized at $1.00$ on every pass, and the two sibling tools on that server remain at $0.00$. A supply-chain change is a step that happens once and then holds, while service-side behavior remains constant, fluctuates within a band or diverges, and divergence is a reason to revalidate a tool against its admission reference, although a refusal still depends on the chosen gate and its calibration.

At $69$ days one more tool, the filesystem server's listing of the directories it may access, moved to $1.00$ because of a change on the measurement host, not in the server. Between the $62$- and $69$-day replays a rename on the host left a symbolic link on the path the server was configured with. The recorded launch command, working directory and environment are identical to $t_0$, and the server now lists the resolved target of that link as a second allowed directory beside the configured one, where it previously listed one. The seven tools of that server that read files and directories stay at $0.00$, and the new listing persists at $77$ days. We exclude this tool and the two harness tools above from every drift count reported in this paper, so the census reports $7$ drifting tools where its raw count at both $69$ and $77$ days is $10$. A more sensitive use-time gate could escalate on this environmental change, while the fixed equality setting below accepts it because its single probe leaves a vacuous threshold.

\paragraph{A fixed gate on the census.} We retrospectively apply the agent study's equality rule at a probe cap of $32$ and $\delta{=}0.05$ to the saved final replay. Each tool's first two admission passes set its reference and self distance, and the current endpoint is the sole candidate. The gate accepts $55/59$ tools and refuses $4/59$ across $416$ probe calls. It refuses three of the seven registry-side changes. Of the four accepted changes, three have just one recorded probe, which makes the radius larger than the maximum possible equality distance. The fourth has a noisy self distance that also makes its threshold vacuous. One overall refusal is a documented CLI timeout artifact. The remaining changes are not adjudicated attacks or independently labeled benign updates, so these decisions measure neither malicious-replacement risk nor a false-alarm rate for legitimate updates.

\section{The agent study and residual failures}
\label{app:agent}

\paragraph{Setup.} MCP-Bench \citep{mcpbench2508} runs an LLM agent over live MCP
servers. We keep the $16$ of its $104$ tasks that touch the four servers for which we hold a gold and a squatter, Math MCP, Unit Converter, Medical Calculator and Scientific Computing. Four further tasks also touch BioMCP, which needs an API key we do not have, and a task whose server cannot connect measures the broken connection instead of the arm, so we drop them and record the omission. Agent and judge are both
Qwen2.5-32B-Instruct-AWQ served at a $65{,}536$ token context, decoding greedily at temperature $0$ with a fixed seed, which removes sampling as a source of difference between the arms without making them identical. The two benign arms recorded $1453$ and $1417$ calls, neither with a corruption, since the serving stack is not reproducible across runs even at temperature $0$, so the arms are paired by task and not by trajectory. Each squatter replaces the gold binary behind an unchanged registry name, so the agent sees byte-identical tool cards, and the arms differ in two environment variables.

\paragraph{Ground truth by gold replay.} Every tool call is recorded with
its arguments and re-issued against the gold server, task by task on freshly
started servers because several of these tools are stateful. A field the gold
server does not reproduce against itself cannot witness a corruption, so each
task is replayed twice and any leaf that differs between the two passes is
excluded and logged. Image responses, which the harness renders as empty text and our parser as a content descriptor, are counted as unadjudicable. The benign arms serve as the known-answer test. After volatile leaves are masked and image responses excluded, every adjudicable benign call reproduces against the gold servers.

\begin{table}[t]
\centering
\caption{\textbf{Verification cuts corrupted calls in four arms over $16$ MCP-Bench tasks}, ground truth by gold-server replay. Corrupted calls are values the agent was given that the gold server would not have returned. With the verifier they fall from $93$ to $20$, and $102$ calls are refused instead, at zero refusals in the benign arm. Each arm was run three times, and the table reports the first, with the spread below.}
\label{tab:agent}
{\small
\setlength{\tabcolsep}{5pt}
\begin{tabular}{lrrrrrr}
\toprule
arm & agent calls & replayed & corrupted & rate & tainted & refused \\
\midrule
benign, no gate    & 1453 & 1388 & 0  & 0.000 & 0/16  & 0 \\
benign, gate       & 1417 & 1347 & 0  & 0.000 & 0/16  & 0 \\
rug pull, no gate  & 1575 & 1476 & 93 & 0.063 & 15/16 & 0 \\
rug pull, gate     & 1323 & 1159 & 20 & 0.017 & 9/16  & 102 \\
\bottomrule
\end{tabular}}
\end{table}

Table~\ref{tab:agent} reports the arms. Taint is counted against the $16$ tasks run, the denominator that makes the gated and ungated arms comparable. The released replay for the gated rug pull carries adjudicable calls for only $15$ of them, because on the unit converter server the gate refused all seven calls the agent made to a server with a gold, so that task has nothing left to adjudicate. Counting taint against those $15$ would report $0.60$ instead of $9/16$ and would drop a task the gate protected completely, so the released replay carries both denominators. The calls a row does not replay are calls to Wikipedia, for which no gold is
held, image responses that cannot be adjudicated, and the gate's refusals,
$65 = 57 + 8$ and $70 = 62 + 8$ in the benign arms, $99 = 89 + 10$ without the gate
and $164 = 52 + 10 + 102$ with it under the rug pull. Replay masked a volatile request
identifier in $2$, $1$, $1$ and $4$ of the $1396$, $1355$, $1486$ and $1169$ calls
sent to it, and the arms ran one after another, not in randomized order. All $102$ refusals fall on the eight tools
whose drift the probes reach, with bmi\_bsa\_calculator $32$,
mean $23$, convert\_temperature $11$, round $11$,
convert\_force $10$, and $5$ each on convert\_density,
convert\_speed and convert\_mass. Verification cost $135{,}111$ probe
calls over the benign run across $269$ verdicts, $502$ per verdict, and produced
no false alarm. Four more verdicts fell on Wikipedia tools, which have no attested
bank and pass through the gate at zero probes, so an average over all
logged verdicts gives $495$.

\paragraph{One coverage failure in two tools.} Ten determinant calls and ten convert\_batch calls remain corrupted under the gate. Both tools sit at a measured distance of $0.000$, not just under a threshold. The determinant tool addresses a named tensor, and a schema-driven probe invents a name that was never created, so gold and squatter both answer that it does not exist. The convert\_batch tool takes a nested array of
request objects and the generator can draw only seven distinct probes for it,
none of which happens to exercise a drifted conversion category. In both cases
the probe distribution does not reach the drifted behavior, so the margin is zero
and Theorem~\ref{thm:bound} gives no detection guarantee at any budget, as the
separability condition of Appendix~\ref{app:sep} states. The budget matters for the
other tools. In a run at $m{=}32$, Unit Converter (realized margin $0.13$ to $0.18$,
requiring $415$ to $774$ probes) and Medical Calculator ($0.093$, requiring
$1576$) were also allowed. Raising $m$ to the full bank of $512$ recovered both,
including the two tools whose budget exceeds $512$, because the budget is
sufficient rather than necessary.

\paragraph{Full-bank use and adaptive exposure.} The gate draws the entire attested bank at each verification. Against an adversary that commits its mimicry region before observing the bank, testing every entry is at least as strong as testing a subset. Full-bank use alone does not invalidate the committed-coverage argument. Reuse creates a different concern, because an adversary that observes the queried inputs can learn the bank and adapt before later checks. That timing is outside the guarantee. A larger bank with withheld probes can delay exposure, but does not establish security against an adversary that learns and adapts to the deployed probes.

\paragraph{Reproducibility of the arms.} Every arm was run three times end to end
on freshly launched servers, and Table~\ref{tab:agent} reports the first. The
agent's trajectory is not identical across runs even under greedy decoding,
because tool responses carry timestamps and the harness retries differently, so
the number of adjudicable calls moves, $1388/1372/1343$ benign ungated,
$1476/1456/1471$ rug pulled ungated, $1347/1361/1365$ benign gated, and
$1159/1154/1104$ rug pulled gated. Task-taint counts do not move. Both benign arms are clean in all six of their runs, with zero corrupted calls and zero refusals every time. The rug pull taints $15$ of the $16$ tasks in all three ungated runs and $9$ in all three gated runs. Corrupted calls track the
trajectory rather than the verdict, $93/90/92$ ungated against $20/16/17$ gated,
a reduction of $78$ to $82\%$ in the count of corrupted calls, with $102/84/77$ calls refused. This records observed run-to-run variation, not a population variance bound, with the agent model, judge and task subset held fixed. The gate's budget and scorer are varied below.

\paragraph{Paired task uncertainty.} In each repeated run, $6$ of the same $16$ tasks change from observed taint to no observed taint and $0$ change in the reverse direction. The paired task-risk reduction is $0.375$, with a $95\%$ task-resampling interval of $[0.125,0.625]$ in each run. Repeats are not additional independent tasks. The interval conditions on the studied servers, and absence of observed corruption does not establish correct final answers. The single-run comparison of equality at $m{=}32$ with structural distance at $m{=}512$ differs by $0.0625$, with interval $[0.0000,0.1875]$. It supports the observed lower probe cost without establishing superior task protection.

\paragraph{Budget and scorer.} Apart from the server's answers, the gate's verdict is a
function of the bank, the budget and the subsample seed, so issuing each full bank once
to the squatter and twice to the gold determines the verdict at every budget and seed
without running the agent. Derived this way, the verdict reproduces the distance,
threshold and probe count the gate logged for every tool in the gated arms, at $m{=}512$
over three seeds and at $m{=}32$. Averaged over $200$ subsample seeds, the structural
distance refuses $2.00$ of the eight drifted tools the probes reach at $m{=}32$, $3.49$ at
$m{=}64$, $6.94$ at $m{=}128$ and $7.98$ at $m{=}256$, and all eight only at the full bank.
Exact equality, calibrated the same way from the two passes over the gold, refuses $7.98$
at $m{=}32$ and all eight from $m{=}64$. Neither refuses a gold tool at any budget. A drifted conversion changes one field of a multi-field response, so equality's margin is
$0.65$ to $0.88$ on the unit conversions and $0.37$ on bmi\_bsa\_calculator, four
to five times the structural distance's $0.13$ to $0.18$ and $0.093$, and the budget of
Theorem~\ref{thm:bound} falls with the square of that ratio. On deterministic tools the
structural distance has a smaller margin because it assigns partial credit, which keeps it continuous under
noise (Appendix~\ref{app:baselines}).

Table~\ref{tab:agentbudget} runs the comparison end to end over the same tasks. The structural
distance leaves $35$ corrupted calls and $13/16$ tainted tasks at $m{=}32$ and again at $m{=}128$,
and only the full bank reaches $20$ and $9/16$. Exact equality at $m{=}32$ leaves $22$ and $8/16$ for
$8{,}412$ probe calls against $126{,}428$, and what survives it in that run is the two tools the probes cannot reach, ten determinant calls and twelve convert\_batch calls. The benign
arm of each configuration is clean, no corrupted call and no refusal, so the lower-cost comparison
detects drift without a false alarm here.

\begin{table}[t]
\centering
\caption{\textbf{Exact equality verifies deterministic drift at a fraction of the probe calls, end to end.} Rug pull arm over the same $16$ tasks, one run each, the $m{=}512$ row being run 1 of Table~\ref{tab:agent}. Corrupted calls and tainted tasks come from the gold replay, refusals from the call log and probe calls from the gate's own log. Every gated row retains the corruption of the stateful tensor tool and of the batch converter, whose probes never reach the drift, and the lower structural budgets retain more.}
\label{tab:agentbudget}
{\small
\begin{tabular}{lrrrrr}
\toprule
per-probe comparison & $m$ & corrupted & tainted & refused & probe calls \\
\midrule
no gate & $0$ & 93 & 15/16 & 0 & 0 \\
structural $d_Y$ & $32$ & 35 & 13/16 & 31 & 8{,}398 \\
structural $d_Y$ & $128$ & 35 & 13/16 & 68 & 33{,}891 \\
structural $d_Y$ & $512$ & 20 & 9/16 & 102 & 126{,}428 \\
exact equality & $32$ & 22 & 8/16 & 84 & 8{,}412 \\
\bottomrule
\end{tabular}}
\end{table}

\paragraph{One fixed replacement setting.} At $m{=}32$ with exact equality, the benign arm allows $278/278$ probed verdicts and makes $10{,}300$ runtime tool calls, including $8{,}846$ probes. The replacement arm allows $237/266$ probed verdicts and makes $9{,}592$ runtime tool calls, including $8{,}412$ probes and $1{,}180$ business calls, while $22/1{,}138$ adjudicable calls remain corrupted. Recording the four trusted banks cost $76{,}174$ calls. The later equality calibration issued another $76{,}174$ trusted calls, so the observed one-time setup cost was $152{,}348$ calls. The benchmark spans $11{,}402$ and $10{,}372$ seconds, respectively, excluding setup, model startup, and gold replay. These arms reused a scorer and budget examined after earlier outcomes, and their agent trajectories differed, so they give a retrospective operating point, not held-out replacement risk.

\paragraph{Limits of this study.} Sixteen tasks are evaluated. Task success is $1.000$ in all four arms, and mean judge-rated completion is $7.750$ and $7.888$ in the benign ungated and gated arms against $7.894$ and $7.969$ in the rug-pull arms. Agent and judge share a
model, because an independent judge from another model family did not follow the rubric's $0$ to $10$ scale and returned per-task
completion scores from $7.9$ to $93.5$. Greedy decoding and a swapped judge mean
none of these scores is comparable to the published MCP-Bench leaderboard, and
every comparison here is paired within the study. The benchmark's own judge is reported only to show what it misses, not as a claim about its quality on the task it was built for.

\section{Random probes from the attested banks}
\label{app:banksample}
The benchmark's cached probes are fixed, so they cannot test the probability statements of Section~\ref{sec:theory}. The banks of the agent study make such a test possible. Its gate recorded every entry of the attested bank of each of $77$ tools on four servers, twice from the gold server and once from the rug-pull server (Appendix~\ref{app:agent}). The tools are stateless on these entries, which Appendix~\ref{app:agent} checks by reproducing every distance the live gate logged, so the recorded per-entry distances determine the empirical distance of any probe subsample and the draws need no further tool calls. For every tool and budget $m\in\{8,16,32,64,128,256,512\}$, with $m$ capped at the bank size, we draw $1{,}000$ subsamples uniformly without replacement and score each draw with both comparisons. The single-candidate gate accepts when the mean distance is at most $s+\epsilon_1$, with $\epsilon_1=\sqrt{\log(2/\delta)/(2m)}$ and, as in the deployed gate, $s$ the admission self distance for the structural distance and the mismatch between the two gold passes for equality. The two-candidate rule of Algorithm~\ref{alg:probeid} uses $\epsilon_2=\sqrt{\log(4/\delta)/(2m)}$, $\gamma_{\min}=2\epsilon_2$ and $\tau_{\mathrm{hi}}=s+\epsilon_2$. Every threshold follows from $\delta{=}0.05$, and the rules, draw seeds and endpoints were fixed before the first draw. A tool counts as wrong when its squatter corrupted at least one adjudicable agent call in an ungated run. Ten tools qualify. The bank reaches the drift of eight of them, and the other two, determinant and convert\_batch, agree with the gold on every bank entry.

The bank is the sampling population, so the bank means $\mu_g$ and $\mu_s$ of the gold and squatter distances are exact expectations under the bank distribution, and each bound's premise can be checked per tool, comparison and budget. Table~\ref{tab:bankcheck} counts the cells where a premise holds and the observed rate falls on the wrong side of its bound. There are none, raw or beyond three Monte Carlo standard errors.

\begin{table}[t]
\centering
\caption{\textbf{Every observed rate respects its bound on the bank distribution.} A cell is one tool, comparison and budget, scored over $1{,}000$ random draws. Premise cells are those where the stated condition holds for the bank means. A violation is an observed rate on the wrong side of the bound.}
\label{tab:bankcheck}
\small
\setlength{\tabcolsep}{4pt}
\begin{tabular}{@{}llcccc@{}}
\toprule
 & & \multicolumn{2}{c}{structural $d_Y$} & \multicolumn{2}{c}{exact equality} \\
\cmidrule(lr){3-4}\cmidrule(lr){5-6}
bound & premise & cells & violations & cells & violations \\
\midrule
replacement refused, rate $\ge1-\delta$ & $\mu_s-s>2\epsilon_1$ & 16 & 0 & 43 & 0 \\
gold accepted, rate $\ge1-\delta$ & $s\ge\mu_g$ & 539 & 0 & 539 & 0 \\
joint wrong decision, rate $\le\delta$ & gold present, $\mu_s>\mu_g$ & 56 & 0 & 56 & 0 \\
gold returned, rate $\ge1-\delta$ & $\mu_s-\mu_g\ge4\epsilon_2$, $s\ge\mu_g$ & 6 & 0 & 27 & 0 \\
\bottomrule
\end{tabular}
\end{table}

\begin{table}[t]
\centering
\caption{\textbf{Acceptance of a replacement with the gold absent}, the fraction of $1{,}000$ random bank draws on which the single-candidate gate accepts the rug-pull squatter, for the ten tools whose squatter corrupted agent calls. Margins are the squatter's bank mean distance. The last two tools match the gold on every bank entry.}
\label{tab:bankaccept}
\small
\setlength{\tabcolsep}{3.5pt}
\begin{tabular}{@{}lcccccccccc@{}}
\toprule
 & \multicolumn{2}{c}{margin} & \multicolumn{4}{c}{structural $d_Y$ at $m$} & \multicolumn{4}{c}{exact equality at $m$} \\
\cmidrule(lr){2-3}\cmidrule(lr){4-7}\cmidrule(lr){8-11}
tool & $d_Y$ & equality & 32 & 128 & 256 & 512 & 8 & 16 & 32 & 64 \\
\midrule
mean & 0.941 & 0.941 & 0.000 & 0.000 & 0.000 & 0.000 & 0.000 & 0.000 & 0.000 & 0.000 \\
round & 0.465 & 0.465 & 0.003 & 0.000 & 0.000 & 0.000 & 0.460 & 0.184 & 0.003 & 0.000 \\
bmi\_bsa\_calculator & 0.093 & 0.371 & 1.000 & 0.997 & 0.060 & 0.000 & 0.651 & 0.435 & 0.046 & 0.000 \\
convert\_density & 0.177 & 0.883 & 1.000 & 0.000 & 0.000 & 0.000 & 0.001 & 0.000 & 0.000 & 0.000 \\
convert\_force & 0.163 & 0.814 & 1.000 & 0.000 & 0.000 & 0.000 & 0.008 & 0.000 & 0.000 & 0.000 \\
convert\_mass & 0.129 & 0.646 & 1.000 & 0.085 & 0.000 & 0.000 & 0.103 & 0.009 & 0.000 & 0.000 \\
convert\_speed & 0.162 & 0.809 & 1.000 & 0.000 & 0.000 & 0.000 & 0.005 & 0.000 & 0.000 & 0.000 \\
convert\_temperature & 0.168 & 0.838 & 1.000 & 0.000 & 0.000 & 0.000 & 0.005 & 0.000 & 0.000 & 0.000 \\
\midrule
determinant & 0.000 & 0.000 & 1.000 & 1.000 & 1.000 & 1.000 & 1.000 & 1.000 & 1.000 & 1.000 \\
convert\_batch & 0.000 & 0.000 & 1.000 & 1.000 & 1.000 & 1.000 & 1.000 & 1.000 & 1.000 & 1.000 \\
\bottomrule
\end{tabular}
\end{table}

Table~\ref{tab:bankaccept} gives the replacement setting tool by tool. Equality's margin on the conversions and on bmi\_bsa\_calculator is four to five times the structural distance's, because a drifted conversion changes one field of a multi-field response, and the refusal budget falls accordingly. Averaged over the eight reachable tools, equality accepts a replacement at $0.154$ at $m{=}8$ and $0.006$ at $m{=}32$ and never from $m{=}64$. The structural distance accepts at $0.750$ at $m{=}32$ and $0.135$ at $m{=}128$, still accepts bmi\_bsa\_calculator at $0.060$ at $m{=}256$, and accepts none at $m{=}512$. Under either comparison, no draw at any budget refuses a gold or any of the $67$ unaffected tools, whose rug-pull server matches the gold on every bank entry and corrupted no agent call, so no benign check is lost at these thresholds. The two unreachable tools pass at every budget. Their margin on the bank distribution is zero, so the theorem makes no claim about them, and their harm lies on the agent's inputs outside the bank.

With the gold present, no draw returns the squatter for any of the ten tools, and on the two unreachable tools every draw abstains because gold and squatter tie. The fraction of draws that return the gold for the eight reachable tools is $0.606$, $0.784$ and $0.934$ for equality at $m{=}16$, $32$ and $64$ and reaches $1.000$ by $m{=}256$. For the structural distance it is $0.250$ from $m{=}128$ to $256$ and $0.750$ at $m{=}512$, because four conversions' margins clear $2\epsilon_2$ only at the full bank, while convert\_mass at $0.129$ and bmi\_bsa\_calculator at $0.093$ never do. These tools are deterministic, so a gold at distance zero cannot be undercut, and the zero wrong-return rate with the gold present follows from that. The informative quantities are the observed replacement acceptance, whose refusal the theory guarantees only where the margin exceeds $2\epsilon_1$, and the absence of false refusals.

A rule that accepts every server is right on the $67$ unaffected tools and wrong on all ten replacements, a tool-level accuracy of $0.870$. In expectation over draws, equality is right on $0.958$ of tools at $m{=}8$ and $0.973$ at $m{=}32$, and the structural distance on $0.890$ and $0.896$, reaching $0.974$ only at $m{=}512$. No gate that accepts every server whose answers match the reference can exceed that value on these banks, because the two unreachable replacements match the gold on every entry. A single-candidate check issues $\min(m,B)$ probe calls and a two-candidate check twice that. Recording the four banks took $76{,}174$ tool calls, and the equality self-distance calibration another $76{,}174$ (Appendix~\ref{app:agent}). The banks were recorded for the agent study, so this is a validation on the bank distributions of four servers rather than on held-out servers, and it says nothing about inputs outside the banks, which is where the two unreachable tools do their harm.

\section{A prospective held-out study}
\label{app:heldout}
The draws of Appendix~\ref{app:banksample} reuse banks that the agent study recorded on servers it had already used, so we also ran a prospective study. Its rules, its endpoints and the code of seven replacement policies, including the salt of their triggers, were fixed before any bank was drawn, and the sharper gates of Appendix~\ref{app:exact} were added only afterwards. The study covers the $38$ read-only tools of eight census servers (Appendix~\ref{app:census}), the MCP reference server, an echo server and the time, memory, SQLite, git, filesystem and weather servers, and leaves out one tool that sleeps by design and one that returns base64 media. Each tool's bank holds $B$ independent draws from the agent study's schema-driven generator, with repeats kept so that the bank is an i.i.d. sample of the input distribution. Paths, time zones, cities, entity and table names and branch types come from typed pools of which $10$ to $30\%$ of draws are invalid or unknown. File paths are drawn from the seven workspace files and four directories, with $15\%$ missing paths and $10\%$ paths outside the workspace. Time zones come from $30$ IANA names and cities from $30$ names, each with $10\%$ invalid or invented values. Memory queries and entity names come from the fixture graph with $30\%$ and $20\%$ random words, table names from the two tables with $30\%$ random names, repository paths with $10\%$ subdirectories and $10\%$ missing paths, and branch types from local, remote and all with $10\%$ random strings. Other fields take the generator's typed values. The bank holds $B{=}512$ calls for the local servers and $96$ for weather, which queries a public weather service at one call per second, and its seed was read from the operating system only after the design was fixed. Every bank was answered three times on freshly started servers, first as the reference, then as the admission pass that fixes the self distance $s$, and last as the later check of the unchanged server, which plays the gold.

Each replacement is one of the fixed policies applied to the gold's answers. Value drift perturbs every scalar leaf of an answer, adding $1$ to $4$ to an integer and $0.5$ to $2$ to a real, appending a suffix to a string and flipping a boolean, on the inputs that a salted hash of the call selects with probability $\phi\in\{1,0.25,0.05\}$. Field drift removes the first field of every mapping answer and falls back to value drift on other answers. Three legitimate updates add a field to every mapping answer, reword error messages, or write integers as floats in reversed key order, which canonicalization removes. The rules and thresholds are those of Appendix~\ref{app:banksample} and Proposition~\ref{prop:sequential} at $\delta{=}0.05$, scored over $1{,}000$ draws per tool. Eight tools have fewer than $16$ distinct inputs in their banks, input-free tools among them, and are reported separately because every draw repeats the same few calls. Of the other $30$, $22$ are local and $8$ belong to the time and weather servers, whose answers depend on the clock or on live weather data. The local tools are deterministic across restarts except for an echo tool whose response carries a timestamp the canonicalizer does not name, which sets its self distance to $0.25$ under $d_Y$ and to $1$ under equality.

No observed rate breaks its bound in any of the $7{,}227$ premise cells, and the two-candidate rules never return a replacement with the gold present (Table~\ref{tab:heldcheck}). The remaining outcomes depend on the margin. Every candidate whose bank margin $\mu-s$ exceeds $0.2$, $117$ tool and candidate cells on $37$ tools under $d_Y$ and $113$ on $33$ under equality, is refused on every draw by the fixed gate at the full bank and by the sequential gate at both resolutions. No gold is refused at the full bank, and at smaller budgets at most $0.1\%$ of gold draws are. On the local tools, full value drift and field drift are refused on all $22$ tools under $d_Y$, and the sequential gate reaches that verdict after $8$ and $14$ probe calls on average, where the fixed gate spends $512$ (Table~\ref{tab:heldpolicy}). Equality accepts every drift policy on the echo tool whose self distance is $1$, because its threshold then admits any candidate, while $d_Y$ refuses them. Drift on one input in twenty is refused on $8$ of the $22$ local tools and accepted on the other $14$, whose $d_Y$ margins lie between $0.004$ and $0.053$, inside the resolution that Proposition~\ref{prop:sequential} leaves open.

\begin{table}[t]
\centering
\caption{\textbf{No observed rate breaks its bound in the held-out study.} Cells are tool, candidate, comparison and budget for the fixed rules, and tool, candidate, comparison and resolution for the sequential rules, over all $38$ tools, the gold and the seven policies. Premises are checked on the bank means. A violation is an observed rate on the wrong side of the bound.}
\label{tab:heldcheck}
\small
\setlength{\tabcolsep}{3.5pt}
\begin{tabular}{@{}lllcccc@{}}
\toprule
 & & & \multicolumn{2}{c}{structural $d_Y$} & \multicolumn{2}{c}{exact equality} \\
\cmidrule(lr){4-5}\cmidrule(lr){6-7}
rule & bound & premise & cells & violations & cells & violations \\
\midrule
fixed & accept, rate $\ge1-\delta$ & $\mu\le s$ & 879 & 0 & 936 & 0 \\
fixed & refuse, rate $\ge1-\delta$ & $\mu-s>2\epsilon_1$ & 586 & 0 & 615 & 0 \\
fixed & joint wrong decision, rate $\le\delta$ & $\mu>\mu_g$ & 1153 & 0 & 1096 & 0 \\
fixed & gold returned, rate $\ge1-\delta$ & $\mu-\mu_g\ge4\epsilon_2$, $s\ge\mu_g$ & 272 & 0 & 313 & 0 \\
sequential & refuse, rate $\le\delta$ & $\mu\le s$ & 262 & 0 & 280 & 0 \\
sequential & accept, rate $\le\delta$ & $\mu>s+\gamma_{\mathrm{res}}$ & 256 & 0 & 242 & 0 \\
sequential & joint wrong decision, rate $\le\delta$ & $\mu>\mu_g$ & 173 & 0 & 164 & 0 \\
\bottomrule
\end{tabular}
\end{table}

\begin{table}[t]
\centering
\caption{\textbf{Held-out outcomes by candidate}, averaged over tools with at least $16$ distinct inputs. Changed counts the tools on which the policy alters at least one answer. Fixed is the acceptance rate of the single-candidate gate at the full bank, sequential its acceptance rate at $\gamma_{\mathrm{res}}{=}0.2$ with the mean number of probe calls per check.}
\label{tab:heldpolicy}
\small
\setlength{\tabcolsep}{3.5pt}
\begin{tabular}{@{}lccccccc@{}}
\toprule
 & & \multicolumn{3}{c}{structural $d_Y$} & \multicolumn{3}{c}{exact equality} \\
\cmidrule(lr){3-5}\cmidrule(lr){6-8}
candidate & changed & fixed & sequential & calls & fixed & sequential & calls \\
\midrule
\multicolumn{8}{@{}l}{\emph{$22$ local tools}} \\
gold (unchanged server) & -- & 1.000 & 1.000 & 128 & 1.000 & 1.000 & 128 \\
value drift, $\phi{=}1$ & 22 & 0.000 & 0.000 & 8 & 0.045 & 0.045 & 14 \\
value drift, $\phi{=}0.25$ & 22 & 0.000 & 0.011 & 82 & 0.045 & 0.056 & 80 \\
value drift, $\phi{=}0.05$ & 22 & 0.636 & 0.744 & 163 & 0.636 & 0.744 & 163 \\
field drift & 22 & 0.000 & 0.000 & 14 & 0.045 & 0.045 & 14 \\
added field & 7 & 0.773 & 0.773 & 115 & 0.818 & 0.818 & 106 \\
reworded errors & 12 & 0.455 & 0.500 & 90 & 0.455 & 0.500 & 90 \\
format only & 0 & 1.000 & 1.000 & 128 & 1.000 & 1.000 & 128 \\
\midrule
\multicolumn{8}{@{}l}{\emph{$8$ time and weather tools}} \\
gold (unchanged server) & -- & 1.000 & 1.000 & 95 & 1.000 & 1.000 & 90 \\
value drift, $\phi{=}1$ & 8 & 0.125 & 0.000 & 23 & 0.250 & 0.125 & 82 \\
value drift, $\phi{=}0.25$ & 8 & 0.500 & 0.126 & 86 & 0.750 & 0.509 & 99 \\
value drift, $\phi{=}0.05$ & 8 & 0.875 & 0.750 & 124 & 0.875 & 0.875 & 107 \\
field drift & 8 & 0.250 & 0.080 & 91 & 0.250 & 0.125 & 82 \\
added field & 6 & 0.375 & 0.250 & 112 & 0.750 & 0.502 & 69 \\
reworded errors & 4 & 0.500 & 0.580 & 118 & 0.500 & 0.625 & 110 \\
format only & 0 & 1.000 & 1.000 & 95 & 1.000 & 1.000 & 90 \\
\bottomrule
\end{tabular}
\end{table}

The time and weather tools show the cost of noise. Five of the six weather tools answer differently on most repeated calls, at self distances up to $0.91$ under $d_Y$, and the remaining margin of a replacement shrinks accordingly. On the noisiest, full value drift changes every answer, yet $s+\epsilon_1$ exceeds the largest possible distance at $B{=}96$, so the fixed gate accepts it on every draw. The sequential gate reports those checks as inconclusive instead, because neither of its tests can pass. A call to the weather service takes a median of $1.6$ to $1.7$ seconds, so a full-bank check of such a tool takes about $4.2$ minutes, while local calls return in about a millisecond and a full-bank check takes $0.66$ seconds.

The legitimate updates measure revalidation cost. The format change leaves every per-entry distance equal to the gold's and is accepted wherever the gold is. Reworded error messages are refused on all $16$ tools whose answers they change. An added field is refused on $5$ of the $7$ local tools it touches under $d_Y$, and passes on two file readers where it alters about one answer in a hundred. Such refusals are the intended behavior of a behavioral reference, which a deployer answers by recording a new one.

The replacements are fixed transformations of genuine answers rather than independent attackers, and the servers are open reference implementations. For the local servers the check followed admission by $22$ to $23$ seconds, so the study does not measure slow drift of an unchanged local server. It tests the stated rules on banks and replacement policies that were fixed in advance, and it does not estimate how often replacements in the wild fall inside the resolution.

\section{Exact and finite-population gates}
\label{app:exact}
The Hoeffding radius behind every gate above is conservative in two ways. It ignores that probes are drawn without replacement from a finite bank, so it stays positive when the whole bank is read, and on the equality channel it ignores that the per-probe score is a Bernoulli indicator. We also score two sharper rules on the same draws and permutations as Appendices~\ref{app:banksample} and~\ref{app:heldout}, at the same $\delta{=}0.05$ without parameter tuning. They were chosen once the results of those appendices were known, so they are reported alongside, without replacing, the original rules. The exact equality gate refuses when $\Pr(X\ge x)\le\delta$, where $x$ counts the mismatching entries among the $m$ drawn and $X$ is hypergeometric with population $B$, $\lfloor sB\rfloor$ successes and $m$ draws. The finite-population radius is $r(m,q)=\sqrt{\rho_m\log(q/\delta)/(2m)}$ with $\rho_m=1-(m-1)/B$ for $m\le B/2$ and $\rho_m=(1-m/B)(1+1/m)$ otherwise \citep[Corollary~2.5]{bardenet2015concentration}, which is zero at $m{=}B$. It uses $q{=}1$ for the fixed single-candidate gate, $q{=}4$ for Algorithm~\ref{alg:probeid} with two candidates, $q{=}2J$ for the single-candidate sequential gate and $q{=}4J$ for the sequential two-candidate rule.

\begin{proposition}[Sharper gates]
\label{prop:exact}
Condition on a fixed bank of per-entry scores and the admission value $s$. If a candidate's bank mean is at most $s$, the exact equality gate refuses it with probability at most $\delta$. With these radii, each single-candidate error statement of Appendix~\ref{app:abstain} holds separately, Proposition~\ref{prop:sequential} holds, and part \emph{(i)} of Theorem~\ref{thm:abstain} holds for two candidates.
\end{proposition}

For a fixed observed count $x$, the hypergeometric upper tail is nondecreasing in the number of mismatching entries, so $\lfloor sB\rfloor$ is least favorable under a bank mean of at most $s$, and the rejection probability is at most $\delta$, often strictly smaller. The proofs of the other statements use only a one-sided deviation bound for each sample mean and a union bound, and Corollary~2.5 of \citet{bardenet2015concentration} provides that bound for sampling without replacement, for the sample mean and, applied to the negated scores, for its lower deviation. $\qquad\blacksquare$

No observed rate breaks its checked bound in any of the $3{,}848$ premise cells on the agent banks or the $8{,}547$ on the held-out study, again an implementation check rather than evidence about the tools. Table~\ref{tab:exact} gives the agent banks. The exact equality gate accepts a reachable replacement at $0.004$ at $m{=}8$ against $0.154$ for the Hoeffding gate, with no recorded acceptance from $m{=}32$ rather than $m{=}64$. It refuses golds and unaffected tools in at most $3.2\%$ of draws, within $\delta$, whereas the Hoeffding gate refuses none. The finite-population radius barely changes the structural distance at small budgets, where $\rho_m$ is close to one, but no recorded draw accepts a reachable replacement from $m{=}256$ rather than $m{=}512$. With the gold present, the finite-population pair rule, which abstains on equal distances as the radius reaches zero, returns the gold on $0.798$ of draws at $m{=}256$ and on every draw at $m{=}512$, against $0.250$ and $0.750$ for the Hoeffding rule.

\begin{table}[t]
\centering
\caption{\textbf{Sharper gates halve the observed refusal budget on the agent banks.} Acceptance of a reachable rug-pull replacement with the gold absent, averaged over the eight reachable tools, and the rate at which the pair rule returns the gold with it present, over $1{,}000$ paired draws per budget. Values are rounded, so a printed zero need not mean zero acceptances. The sharper rules were added after the Hoeffding results were known.}
\label{tab:exact}
\small
\setlength{\tabcolsep}{4pt}
\begin{tabular}{@{}rcccccc@{}}
\toprule
 & \multicolumn{2}{c}{equality, gold absent} & \multicolumn{2}{c}{structural $d_Y$, gold absent} & \multicolumn{2}{c}{$d_Y$, gold returned} \\
\cmidrule(lr){2-3}\cmidrule(lr){4-5}\cmidrule(lr){6-7}
$m$ & Hoeffding & exact & Hoeffding & finite pop. & Hoeffding & finite pop. \\
\midrule
8 & 0.154 & 0.004 & 0.807 & 0.807 & 0.000 & 0.000 \\
16 & 0.079 & 0.000 & 0.773 & 0.760 & 0.128 & 0.128 \\
32 & 0.006 & 0.000 & 0.750 & 0.750 & 0.158 & 0.158 \\
64 & 0.000 & 0.000 & 0.552 & 0.238 & 0.244 & 0.247 \\
128 & 0.000 & 0.000 & 0.135 & 0.075 & 0.250 & 0.250 \\
256 & 0.000 & 0.000 & 0.007 & 0.000 & 0.250 & 0.798 \\
512 & 0.000 & 0.000 & 0.000 & 0.000 & 0.750 & 1.000 \\
\bottomrule
\end{tabular}
\end{table}

On the held-out study the full bank is read, so the finite-population radius is zero and the gate refuses any candidate whose bank mean exceeds $s$. This removes sampling error within the bank but says nothing about inputs outside it. Every tested malicious replacement with a positive bank margin is then refused, including drift on one input in twenty, which the Hoeffding gate accepted on $14$ of the $22$ local tools, and full value drift on the noisiest weather tool. No gold is refused at the full bank under either sharper rule, and at smaller budgets at most $4.7\%$ of gold draws are refused by the exact gate and $0.8\%$ under the finite-population radius. The cost is refusal of legitimate updates. At the full bank the finite-population structural gate refuses every answer-changing legitimate update tested, including an added field on all $7$ local tools it touches rather than $5$. The sharper rules also make the admission estimate $s$ binding. Reading the whole bank, the gate refuses a gold whose bank mean exceeds that estimate by any amount, which none of the held-out golds did, but their checks followed admission by minutes at most.

\section{Implementation and reproducibility}
\label{app:impl}
\begin{table}[t]
\centering
\caption{\textbf{Study conclusions depend on the decision rule.} Forced choice treats a gold tie as a miss. Gated selection uses Algorithm~\ref{alg:probeid}. The single-candidate gate accepts distance at most $\tau_{\mathrm{hi}}$ (Appendix~\ref{app:abstain}). The gated thresholds $0.5$ and $0.05$ are fixed defaults of the released configuration, not derived from the theory, and the twin family uses $0.6$. Appendix~\ref{app:riskcov} scores the theory-calibrated radius on the cached cases. The sequential gate also accepts once $d^{(j)}+\epsilon_j\le s+\gamma_{\mathrm{res}}$, and its paired form uses the gap $2\epsilon_j$.}
\label{tab:rules}
\small
\setlength{\tabcolsep}{4pt}
\begin{tabular}{@{}p{0.50\textwidth}lcc@{}}
\toprule
study & rule & $\tau_{\mathrm{hi}}$ & $\gamma_{\min}$ \\
\midrule
Tables~\ref{tab:main}, \ref{tab:judge} and~\ref{tab:judgefull} (ProbeID and judge view; the judges abstain on 0), \ref{tab:baselines}, \ref{tab:probegen}, \ref{tab:multiserver}, \ref{tab:reimpl} (sloppy and faithful); Figures~\ref{fig:main}(c), \ref{fig:descsim}, \ref{fig:realbudget}; Appendices~\ref{app:canon}, \ref{app:ksweep} & forced choice & -- & -- \\
Tables~\ref{tab:adaptivesweep}, \ref{tab:llmsquatter}, \ref{tab:natural}, \ref{tab:jsonimpl}, \ref{tab:reimpl} (twin), \ref{tab:stateful}, \ref{tab:sideeffect}; sibling groups & gated & $0.5$ & $0.05$ \\
Table~\ref{tab:twin} & gated & $0.6$ & $0.05$ \\
Table~\ref{tab:dilution} & margin $\ge\gamma_{\min}$ & -- & $0.05$ \\
Appendix~\ref{app:riskcov} & calibrated, swept & -- & -- \\
Tables~\ref{tab:agent}, \ref{tab:agentbudget}; census gate (Appendix~\ref{app:census}) & one-sided gate & self dist.\ $+\,\epsilon$ & -- \\
Table~\ref{tab:operating} (bank rows, gold absent or alone), \ref{tab:bankaccept}, \ref{tab:bankcheck}, \ref{tab:heldcheck}, \ref{tab:heldpolicy} (fixed); Figure~\ref{fig:main}(d) & one-sided gate & $s+\epsilon_1$ & -- \\
Table~\ref{tab:operating} (bank row, gold present), \ref{tab:bankcheck}, \ref{tab:heldcheck} & Algorithm~\ref{alg:probeid} & $s+\epsilon_2$ & $2\epsilon_2$ \\
Tables~\ref{tab:heldcheck}, \ref{tab:heldpolicy} (sequential); Proposition~\ref{prop:sequential} & sequential gate & $s+\epsilon_j$ & -- \\
Table~\ref{tab:exact} (equality, gold absent); Appendix~\ref{app:exact} & exact test & hypergeometric & -- \\
Table~\ref{tab:exact} ($d_Y$, gold absent) & one-sided gate & $s+r$ & -- \\
Table~\ref{tab:exact} ($d_Y$, gold present) & pair rule & $s+r$ & $2r$ \\
\bottomrule
\end{tabular}
\end{table}

On the cached benchmark, identity verification is inference only and deterministic given the saved responses and the probe seed. The benchmark backend is the StableToolBench response cache. The recursive loader admits a tool when it exposes at least four structured cached responses, parsing argument keys with a safe literal evaluator and skipping binary or empty payloads, and the pool of Table~\ref{tab:main} contains every admitted tool matched to a ToolBench description, $402$ tools. For each registry seed, $41$, $42$ and $43$, the pool is shuffled and golds are taken in that order until $150$ are kept, $300$ for Appendices~\ref{app:judge} and~\ref{app:baselines}. A case is kept only when at least one of its squatters is behaviorally separable from the gold, so hit@1 is conditional on that event. Reaching $150$ golds per seed drew $462$ golds over the three seeds at $k{=}1$ and $451$ at every $k$ from $2$ to $64$, so $12$ and $1$ were excluded. Every gold excluded at $k{=}1$ returns canonical responses holding no value a mutation changes, and every gold kept holds one. On this 150-gold cohort, counting the excluded golds as misses lowers ProbeID from $1.000$ to $0.974$ at $k{=}1$ and from $0.956$ to $0.953$ for $k\ge2$. The $300$-gold studies drew $929$ golds at $k{=}1$ and $903$ for $k\ge2$. A gold's probes are its first $12$ cached argument and response pairs in cache order, fewer when the cache holds fewer. Squatter $j$ takes the family at position $j$ of the ordered pair (output value drift, silent field drift). Output value drift perturbs every scalar leaf, negating a boolean, adding a uniform integer from $1$ to $4$ or a uniform real from $0.5$ to $2.0$, and appending a suffix to a string. Silent field drift deletes the first key of a mapping response in sorted order, and falls back to output value drift when the response is not a nonempty mapping. Mutations, probe order and seeds are fixed in code, the gated studies use the code's default thresholds, no development split was held out, and Table~\ref{tab:rules} lists the decision rule behind each study. The cached identity results average three registry seeds, and a gold can recur across them. At $k{=}8$ the $450$ case-runs of Table~\ref{tab:main} are $307$ distinct golds, $99$ of them in two seeds and $22$ in all three. Resampling golds, each with all its case-runs, instead of case-runs leaves the gap over BM25 at $0.844$ and moves its $95\%$ interval from $[0.809, 0.878]$ to $[0.807, 0.880]$. The bounded behavioral distance, the identity rule, and the guarantee check are released with the benchmark, together with the per-case hit vectors and the paired bootstrap output behind the headline confidence intervals. The live study uses a stdio MCP client backend that implements the same interface as the cached backend, with the official everything server as gold and a small squatter server, both released. The ProbeID code is released under the MIT license, and the benchmark builds on StableToolBench and ToolBench, both under the Apache-2.0 license. Code, derived data and the full raw response corpus will be released upon acceptance.

\paragraph{Cost.} The protocol reads the reference from its attested bank. With the full budget $m{=}12$ and one gold plus $k$ squatters, $m(k{+}1)$ is a cap of $24$, $60$, and $108$ candidate calls at $k\in\{1,4,8\}$. Fewer cached probes lower the observed number of candidate-response comparisons. The saved $k{=}8$ cases average $87.64$ comparisons, which are not measured live calls. A variant that also queried the trusted reference live would add another $m$ calls. The behavioral judge receives the same probes with each response cut to $200$ characters and makes one LLM request per selection, while description routers issue no selection-time tool calls. The cached benchmark sweep, covering 150 gold, $k\in\{1,4,8\}$, three seeds, and four selectors, takes $18.8$ seconds on one CPU core. That wall time excludes live tool latency and cannot establish end-to-end speed. Per-case distances and threshold sweeps are retained alongside the benchmark. The deterministic dilution cells have no trial denominator.

\end{document}